\documentclass[%
 reprint,
nofootinbib,
 amsmath,amssymb,
 aps,
]{revtex4-2}
\usepackage{textcomp}
\usepackage{graphicx}
\usepackage{xcolor}
\usepackage{dcolumn}
\usepackage{bm}
\usepackage{hyperref}
\usepackage{cleveref}
\usepackage{siunitx} 

\newcommand{\dd}{\text{d}}

\DeclareSIUnit\Molar{M}

\begin{document}

\title{Extension of the primitive model by hydration shells and its impact on the reversible heat production during the buildup of the electric double layer}

\author{Philipp Pelagejcev}
\email{philipp.pelagejcev@physik.uni-freiburg.de}

\author{Fabian Glatzel}
\email{fabian.glatzel@physik.uni-freiburg.de}

\author{Andreas H\"artel}%
\email{andreas.haertel@physik.uni-freiburg.de}
 
\affiliation{%
 Institute of Physics, University of Freiburg, Hermann-Herder-Str. 3, 79104 Freiburg, Germany
}%

\date{\today}

\begin{abstract}
Recently the reversible heat production during the electric double layer (EDL) buildup in a sodium chloride solution was measured experimentally [Janssen \emph{et al.,} Phys. Rev. Lett. \textbf{119}, 166002 (2017)] and matched with theoretical predictions from density functional theory and molecular dynamics simulations [Glatzel \emph{et al.,} J. Chem. Phys. \textbf{154}, 064901 (2021)]. In the latter, it was found that steric interactions of ions with the electrode’s walls, which result in the so-called Stern layer, are sufficient to explain the experimental results. As only symmetric ion sizes in a restricted primitive model were examined, it is instructive to investigate systems of unequal ion sizes that lead to modified Stern layers. In this work, we explore the impact of ion asymmetry on the reversible heat production for each electrode separately. In this context, we further study an extension of the primitive model where hydration shells of ions can evade in the vicinity of electrode’s walls. We find a strong dependence on system parameters such as the particle sizes and the total volume taken by particles. Here, we even found situations where one electrode was heated and the other electrode was cooled at the same time during charging, while, in sum, both electrodes together behaved very similarly to the already mentioned experimental results. 
Thus, heat production should also be measured in experiments for each electrode separately. By this, the importance of certain ingredients that we proposed to model electrolytes could be confirmed or ruled out experimentally, finally leading to a deeper understanding of the physics of EDLs.\\
\\
\textbf{Published:} J. Chem. Phys. \textbf{156}, 034901 (2022); \url{https://doi.org/10.1063/5.0077526}

\end{abstract}

\maketitle

\section{Introduction}
The reversible heat involved in the buildup of electric double layers (EDLs) has been studied in recent works experimentally \cite{schiffer2006heat,janssen2017coulometry} and theoretically \cite{Theodoor1990role, Biesheuvel2007counterion, dEntremont2014first, dEntremont2015thermal, janssen2017reversible, cruz2018electrical, alizadeh2020temperature,glatzel2021reversible}. In one of these studies it has been claimed that heat effects in EDLs are described theoretically by simply modeling the Stern layer separation correctly \cite{glatzel2021reversible}. As a consequence, first, rather simple models would be sufficient to describe heat effects in capacitors or biological systems. Second, huge anisotropies would be expected for EDLs in electrolytes whose constituents are of different sizes and shapes.

EDLs are established by mobile charges at (charged) interfaces. In the context of technical applications, EDLs appear in supercapacitors. The latter store electric energy \cite{Beidaghi2014capacitive} but further can be used in purifying (desalinating) water \cite{Suss2015water, Kim2015enhanced} and harvesting energy from concentration (the so-called blue energy) and thermal gradients \cite{Brogioli2009extracting, Jia2014blue, Janssen2014boosting, Haertel2015heat}. Here, thermal properties of these systems are crucial to ensure optimal and efficient operation and enhanced longevity of its components. As another example in a completely different context, the relation of heating and charging is also relevant for nervous conduction in biology and medicine \cite{Shapiro2012Infrared, Plaksin2018Thermal,Lichtervelde2020heat}. Hence, theoretical understanding of EDLs is essential. 

A common approach to describe EDLs is the Gouy-Chapman-Stern model, where the EDL is modelled within Poisson-Boltzmann theory by a composition of point charges that form a so-called diffusive layer \cite{Gouy1910sur, Chapman1913contribution}. An additional Stern layer \cite{Stern1924theorie} that limits the minimal distance between the ions in the electrolyte and the electrode can further capture key properties due to finite ionic volume. This concept is naturally included in the restricted primitive model (RPM), where the electrolyte is modelled by charged hard spheres. The RPM and, in particular, its structural properties are well described in the theoretical framework of classical density functional theory (DFT) \cite{Evans1979nature}, because the framework remarkably well handles hard-sphere interactions within fundamental measure theory \cite{Rosenfeld1989free, Roth2010fundamental, Oettel2012description, Haertel2015anisotropic}. In consequence, DFT is able to describe the structure of the first layers of ions in the vicinity of an electrode \cite{Roth2016shells, Haertel2017structure, Cats2021primitive}, a property that has recently been discussed to be key for properties such as the heat production during charging \cite{glatzel2021reversible} and the differential capacitance in EDLs \cite{Cats2021differential}. 

The mobile charges are typically immersed in a surrounding solvent. This solvent is usually modelled implicitly by a homogeneous dielectric background, partly reflecting that explicit dipolar solvent particles would arrange around ions thereby forming hydration shells and, as a consequence, influencing the structure of the electrolyte and the local permittivity \cite{Tansel2006significance, Lyashchenko2010dielectric, Bankura2013hydration, nightingale1959phenomenological}. Likewise, dipolar solvents as water would show interfacial solvent (hydration shell) depletion that would lead to a non-local dielectric permittivity in the vicinity of surfaces \cite{henderson2005monte,buyukdagli2012dipolar, buyukdagli2012excluded}. Consequently, during charging a capacitor, the solvent would be repelled from the EDL region and ions in this region would become dehydrated, effectively resulting in a decrease of the Stern layer separation. We call this phenomenon hydration-shell evasion (HSE) and propose a simplistic extension of the (restricted) primitive model to account for this effect. For this purpose, we describe the hard interaction of each ion not only by a 
hydrated diameter $d^{\mathrm{hyd}}$, which leads to steric repulsion among all ion species, but also by a crystallized diameter $d^{\mathrm{cry}} < d^{\mathrm{hyd}}$, which governs the steric repulsion between an individual ion species with the electrodes. Note that each ion species is defined by a set of two diameters which can vary for different ion species, thus, there are more possibilities to construct size-asymmetric electrolytes. Due to size asymmetry in the ion species the electrodes must be studied separately.

In this work, our observable of choice is the reversible heat because it was found to strongly depend on the Stern layer \cite{glatzel2021reversible} and, therefore, on the crystallized instead of the hydrated diameter. We first briefly recapitulate the experiment in which the reversible heat production during the EDL buildup has been measured (in sum over both electrodes) \cite{janssen2017coulometry} and its theoretical description in DFT \cite{glatzel2021reversible}. Second, we extend the RPM to describe hydration-shell evasion within DFT. Then, starting from the RPM, we gradually increase the complexity of the model electrolyte and study the effects of hydration-shell evasion on the EDL. We, additionally, discuss dehydration energies, supported by another simple model extension. We conclude with an outlook on potential future experiments beyond those of ref.~\cite{janssen2017coulometry} and its value for our general understanding of electrolytes and EDLs. 

\section{Model}
\subsection{Experimental setup and thermodynamic considerations}
\begin{figure*}
	\includegraphics[width=\linewidth]{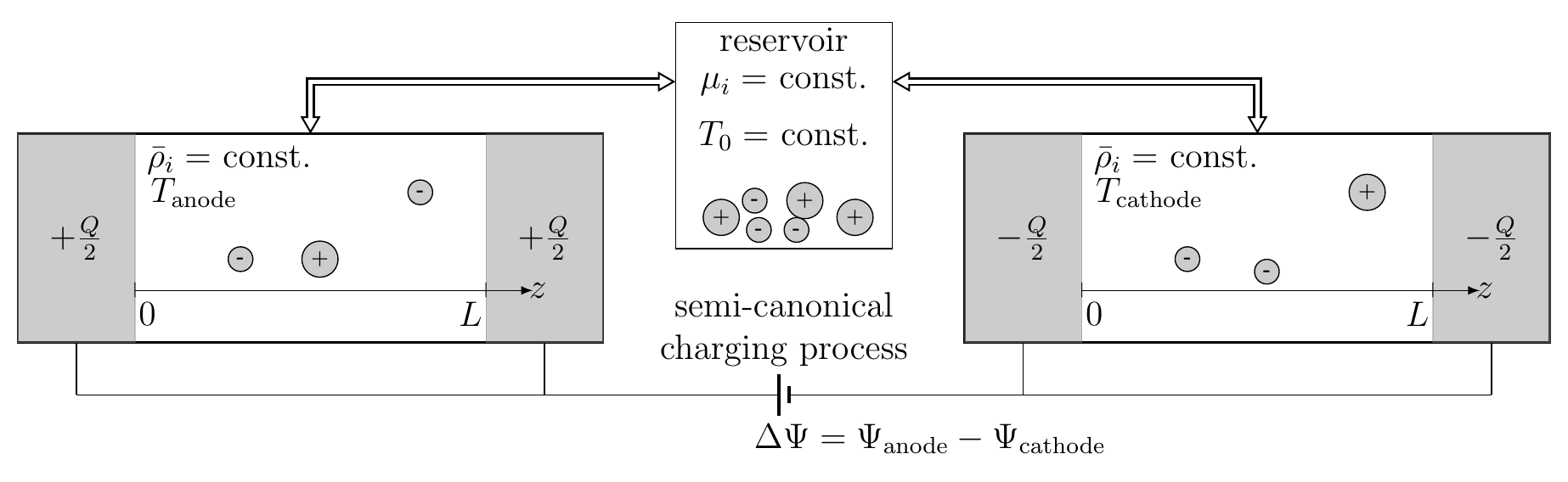}
	\caption{Illustration of the model system used to mimic the experimental setup of ref.~\cite{janssen2017coulometry}. Here, two separate subsystems with planar symmetry are used to model the two differently charged (porous) electrodes. Both are coupled to a common particle and heat reservoir fixing the chemical potentials and the temperature during the charging process. Accordingly, the total number of ions in the subsystems may change during (dis)charging. Each subsystem consists of two equally charged parallel hard walls. Depending on the distance $L$ between those walls, EDLs of the same kind may overlap. The same effect may occur in smaller pores of the porous carbon electrodes in the experiments reported in ref.~\cite{janssen2017coulometry}.}\label{Fig:SemiCanonical}
\end{figure*}
In the experimental setup of ref.~\cite{janssen2017coulometry}, two nanoporous carbon electrodes were immersed into a larger container filled with a sodium chloride solution, which acted as a particle bath for the system. A temperature probe between the electrodes measured the increase/decrease $\Delta T$ of temperature $T$ during (dis)charging from which the amount of reversible heat $\mathcal{Q}_\textrm{rev}$ produced in sum at both electrodes was determined. In the experiment, the reversible increase/decrease in temperature was comparably small ($\Delta T \ll \SI{1}{K}$) and depended on the heat capacity of the system. Considering this, we assume an isothermal (dis)charging process in the following. Thus, the whole calculation does become independent of the heat capacity.

The pores in the electrodes used in the experimental setup have various shapes and sizes.  
In order to rule out effects from different geometries in our work that, of course, could be important, we study EDLs at perfectly flat parallel walls; once effects are clear in this setting, extensions and effects from additional geometries can be studied. 

We describe one nanopore in a nanoporous electrode as two equally charged planar hard walls with distance $L=\SI{10}{\nm}$ (if not mentioned otherwise) and a surface area $A$ per wall as done in ref.~\cite{glatzel2021reversible}. The setup is sketched in Fig.~\ref{Fig:SemiCanonical}. This distance is chosen such that the EDLs do not overlap which would be the case particularly in small pores.
The electrodes carry a surface charge density $\pm e\sigma=\pm Q/(2A)$ due to the applied surface potential $\Psi$, where $e$ is the proton charge. We use $\Psi=\psi(z=0)=\psi(z=L)$ to explicitly refer to the surface potential at the electrode. Consequently, $\Delta\Psi$ is the potential difference between both electrodes. We assume ideal walls of infinite extension such that translational symmetry along the wall allows us to describe positions in the system by an effectively one-dimensional spatial coordinate $z$ perpendicular to the walls. In between those walls, the electrolyte is represented in the (solvent) primitive model, which will be discussed in Sec.~\ref{sec:IonModels}.

As described in ref.~\cite{glatzel2021reversible}, the electric work $\Delta W_\textrm{el}$ for isothermal charging of one electrode from surface charge $Q_1$ to surface charge $Q_2$ is given by
\begin{equation}
	W_\mathrm{el} = \int_{Q_1}^{Q_2} \Psi \mathrm{d}Q^\prime = \Delta \Omega,
\end{equation}
where $\Delta\Omega$ is the change in the grand potential. Equally, the reversible heat can be expressed via the entropy difference $\mathcal{Q}_\mathrm{rev}=T \Delta S$. Intuitively, the heat flow at an electrode can be thought of in the following manner: Charging the electrode orders the solvent at the electrode, hence decreasing the entropy of the system, which is accompanied by a temperature change. However, as discussed in ref.~\cite{glatzel2021reversible}, the amount of reversible heat produced during charging sensitively depends on the chosen ensemble. It turns out that fixing the bulk densities $\bar{\rho}_i$ is best suited for the description of the experiments in ref.~\cite{janssen2017coulometry}. By doing so, neither do the bulk densities $\bar{\rho}_i$ (set by chemical potentials $\mu_i$) change during (dis)charging (similar to a grand canonical description) nor do the bulk densities change (the average number of particles per volume) with an increasing/decreasing temperature (similar to a canonical description). In particular, this implies that the chemical potentials change with temperature ensuring constancy of the bulk densities. As in ref.~\cite{glatzel2021reversible}, we refer to such a description as \emph{semi-canonical}; see Fig.~\ref{Fig:SemiCanonical}. Employing this description, the change in entropy can be determined as
\begin{equation}\label{equ:Entropy}
	\Delta S = \left( \frac{\partial \Omega}{\partial T} \right)_{\bar{\rho}_i, Q_2} - \left( \frac{\partial \Omega}{\partial T} \right)_{\bar{\rho}_i, Q_1},
\end{equation}
where the derivatives are taken at fixed bulk densities and surface charges. Accordingly, the heat during an isothermal charging process is as follows
\begin{align}
\mathcal{Q}_\text{rev} = T \int_{Q_1}^{Q_2} \left(\frac{\partial S}{\partial Q'}\right)_{T,\bar{\rho}_i} dQ' . 
\end{align}

\subsection{Ion models}
\label{sec:IonModels}
\begin{figure}
	\includegraphics[width=\linewidth]{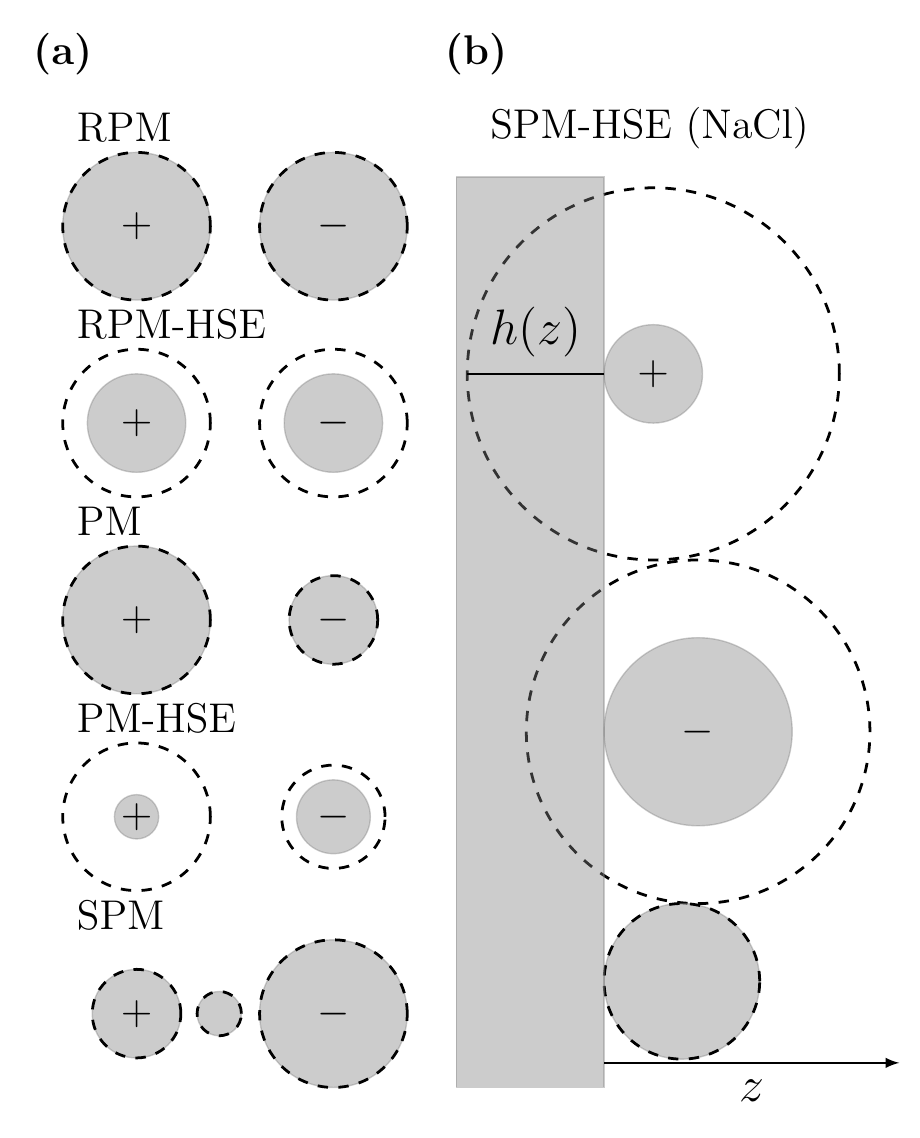}
	\caption{\textbf{(a)} Schematic representation of the different ion models. The hydrated ion diameter is depicted as a white circle with the dashed border line, and the crystallized diameter is depicted as a gray core circle within the ion. \textbf{(b)} Up to scale representation of the SPM-HSE with diameter values for sodium chloride from ref.~\cite{nightingale1959phenomenological}. The hydrated part of an ion can permeate the electrode (grey rectangle on the left). The permeated volume of an ion at position $z$ is a spherical cap with height $h(z)$, depicted as a line.}\label{Fig:HSE}
\end{figure}
In this work, we study different models describing the EDL buildup by ions and solvents.
We model the electrolyte via the (restricted) primitive model, (R)PM, where each particle $i$ is modeled as a hard sphere with diameter $d_i$ (in the RPM all diameters are the same). Furthermore, ions carry an additional point charge. We consider only monovalent ions such that the valency $z_+=-z_-=1$, where the 
indices $+$/$-$ label cations and anions, respectively. The solvent is modeled implicitly as a dielectric background with relative permittivity $\epsilon_r=80$ to represent water. When we consider explicit steric ion-solvent interactions, we additionally introduce neutral solvent particles with $z_\circ=0$. Hence, the dimensionless interaction in the (R)PM between two particles $i$ and $j$ is given by

\begin{equation}\label{eq:PM_Interaction}
\beta \phi_{ij}^{\textrm{(R)PM}}(r_{ij}) =
\begin{cases}
\infty, & r_{ij} < \frac{1}{2} (d_i+d_j) \\
\lambda_{\textrm{B}}\frac{z_i z_j}{r_{ij}}, & r_{ij} \geq \frac{1}{2} (d_i+d_j) , 
\end{cases}
\end{equation}
where $\beta=1/k_\mathrm{B}T$ is an inverse temperature with the Boltzmann constant $k_\mathrm{B}$, $r_{ij}$ is the distance between the particles $i$ and $j$, and $\lambda_{\textrm{B}}= e^2/(4\pi \epsilon_0 \epsilon_r k_{\textrm{B}}T)$ is the Bjerrum length that contains the vacuum permittivity $\epsilon_0$.

The electrodes interact with the particles in two ways. First, we capture their hard repulsion in the external potential
\begin{equation}\label{eq:Ext_Potential}
    V^{\textrm{ext}}_i(z) =
    \begin{cases}
    \infty, & z < \frac{1}{2}d_i \text{ or } L-z < \frac{1}{2}d_i \\
    0, & \text{else}
    \end{cases}
\end{equation}
that acts on each particle $i$. The electrostatic interaction due to surface charges at the electrodes is considered in the total charge distribution that enters the mean-field electrostatic functional and adds a boundary condition to the Poisson equation, as we discuss within the next Sec.~\ref{sec:DFT}.

In particular, we study hydration-shell evasion (HSE) in the direct vicinity of the solid electrode via an extension of the (R)PM.
For this purpose, we distinguish between the diameters that enter the particle interaction potential in eq.~(\ref{eq:PM_Interaction}) and the external potential in eq.~(\ref{eq:Ext_Potential}), as sketched in Fig.~\ref{Fig:HSE}: While the particle interaction potential mainly models the bulk behavior and, for this reason,  the contributing diameter $d^\text{hyd}$ is the hydrated one which explicitly contains hydration shells, we use a smaller diameter $d^\text{cry}$ for the external potential that rather corresponds to the bare crystalline ion diameter without hydration shell. Hence in the HSE model extension each particle species $i$ has a set of two different diameters, $d^\text{hyd}_i$, that determines the particle interaction potentials in eq.~(\ref{eq:PM_Interaction}), and $d^\text{cry}_i$, that dictates the onset of the external potential in eq.~(\ref{eq:Ext_Potential}). For the interpretation of these diameters as hydration-shell evasion in the vicinity of the electrode one has to set $d_i^{\textrm{cry}} < d_i^\textrm{hyd}$.
Consequently, the Stern layer, which covers the key physics of reversible heat production \cite{glatzel2021reversible}, is mainly determined by the crystallized diameter $d^\text{cry}$. Note that the (R)PM is recovered for $d_i^\text{hyd}=d_i^\text{cry}$.

Several ion models accrue from this HSE model extension: In the case for the restricted primitive model with hydration shell evasions (RPM-HSE) the corresponding ion diameters are equal, i.e. $d^{\mathrm{hyd}}_{+}=d^{\mathrm{hyd}}_{-}=d^{\mathrm{hyd}}$ and $d^{\mathrm{cry}}_{+}=d^{\mathrm{cry}}_{-}=d^{\mathrm{cry}}$. By additionally setting $d^{\mathrm{cry}} = d^{\mathrm{hyd}}$ one retrieves the RPM. Otherwise when the ion diameters, hydrated or crystallized, differ between the ion species, we refer to it as primitive model with hydration-shell evasion (PM-HSE). For unequal crystallized diameters the Stern layer is affected by all these ion diameters. Hence, the reversible heat production in this case should be qualitatively different from cases with symmetric diameters \cite{valleau1982electrical, yu2006effects}. 

In the solvent primitive model (SPM) steric interactions of the solvent are modeled explicitly by an additional neutral hard-sphere species, which, of course, affects the Stern layer. The neutral solvent hard spheres do not have a hydration shell, hence they are defined by a single diameter $d_{\circ}$.

To be able to compare with experimental results from ref.~\cite{janssen2017coulometry}, where a sodium chloride solution was used, we choose in the PM-HSE and SPM-HSE the sodium diameters as $d^{\mathrm{hyd}}_{+}=\SI{0.716}{\nm}$ and $d^{\mathrm{cry}}_{+}=\SI{0.19}{\nm}$, as well as the chloride diameters $d^{\mathrm{hyd}}_{-}=\SI{0.664}{nm}$ and $d^{\mathrm{cry}}_{-}=\SI{0.362}{nm}$. These effective values are taken from scattering measurements, reported in ref.~\cite{nightingale1959phenomenological}. This is an especially interesting case, because $d^{\mathrm{hyd}}_{+} > d^{\mathrm{hyd}}_{-}$ but $d^{\mathrm{cry}}_{+} < d^{\mathrm{cry}}_{-}$. 
The diameter of the neutral solvent particles is set to $d_{\circ}=\SI{0.3}{\nm}$ in order to be able to compare to previous work \cite{glatzel2021reversible} and $d^{\mathrm{cry}}_{+}<d_{\circ}<d^{\mathrm{cry}}_{-}$. Again, setting equal values for crystallized and hydrated diameters results in SPM. In the case of SPM (NaCl) we also use the hydrated diameters for the steric ion-wall interaction. 
All discussed ion models are depicted in Fig.~\ref{Fig:HSE} and our commonly used numerical diameter values are summarized in Table~\ref{Tab:IonModels}. 

An important measure for systems of hard particles is the volume fraction. The total bulk volume fraction of the system is defined by
\begin{equation}
\eta=\sum_{i \in \{+,-\}}\frac{\pi}{6}\bar{\rho}_i (d^\mathrm{hyd}_i)^3 + \frac{\pi}{6}\bar{\rho}_\circ d_\circ^3
\end{equation} 
which is the fraction of the total occupied volume by all particles divided by the total available system volume. In cases where we fix the volume fraction we choose $\eta\approx 0.468$, a value just below $0.49$ where the fluid-solid phase transition in monodisperse hard spheres would occur \cite{Hoover1968melting, Oettel2010free}. 

\begin{table}
	\renewcommand{\arraystretch}{1.3}
	\begin{tabular}{|l|c|c|c|c|c|}
		\hline
		ion model & $d_{+}^{\mathrm{hyd}}$ & $d_{-}^{\mathrm{hyd}}$ &  $d_{+}^{\mathrm{cry}}$ &
		$d_{-}^{\mathrm{cry}}$ & $d_\circ$ \\
		\hline 
		RPM & \multicolumn{2}{c|}{$\SI{0.68}{\nm}$} & \multicolumn{2}{c|}{$\SI{0.68}{\nm}$} &  $\diagup$ \\
		\hline
		RPM-HSE & \multicolumn{2}{c|}{$\SI{0.68}{\nm}$}  & \multicolumn{2}{c|}{variable} & $\diagup$ \\
		\hline
		PM &   $\SI{0.716}{\nm}$ &  $\SI{0.664}{\nm}$ & $\SI{0.716}{\nm}$ &  $\SI{0.664}{\nm}$ & $\diagup$ \\
		\hline
		PM-HSE & $\SI{0.716}{\nm}$ & $\SI{0.664}{\nm}$ &$\SI{0.19}{\nm}$ &$\SI{0.362}{\nm}$ & $\diagup$ \\
		\hline
		SPM &   $\SI{0.716}{\nm}$ &$\SI{0.664}{\nm}$ &  $\SI{0.716}{\nm}$ &$\SI{0.664}{\nm}$ &$\SI{0.3}{\nm}$ \\
		\hline
		SPM-HSE & $\SI{0.716}{\nm}$ & $\SI{0.664}{\nm}$ & $\SI{0.19}{\nm}$ & $\SI{0.362}{\nm}$ & $\SI{0.3}{\nm}$ \\
		\hline
	\end{tabular}
	\caption{Overview of the ion models defined in this work and of the often used numerical values for diameters therein. If not mentioned otherwise, we use these values.}\label{Tab:IonModels}
\end{table}

\subsection{Representation of the HSE model extension in DFT}\label{sec:DFT}
In classical density functional theory (DFT) the object of interest is a functional $\Omega[\{\rho_i\}]$ of the one-particle densities $\rho_i$. This functional can be minimized in order to obtain the equilibrium density profiles $\rho_i^{\textrm{eq}}$ and the respective grand canonical potential energy $\Omega[\{\rho_i^{\textrm{eq}}\}]$. While the existence of such an exact functional has been proven \cite{Hohenberg1964inhomogeneous, Mermin1965thermal,Evans1979nature}, its approximation is typically tricky. Nonetheless, once a well-performing functional has been found, all thermodynamic properties are available via the grand potential that follows from a variational principle and, thus, the minimization
\begin{equation}
\left. \frac{\delta\Omega[\{\rho_i\}]}{\delta\rho_j} \right|_{\rho_i=\rho_i^{\textrm{eq}}}=0  \quad \forall \text{ species } j
\end{equation}
of the functional. Consequently, the reversible heat $\mathcal{Q}_\textrm{rev}=T\Delta S$ can be obtained directly in the framework of DFT via eq.~(\ref{equ:Entropy}), since one obtains the grand potential $\Omega(T,V,\mu, Q)$ by evaluating the functional for the minimizing equilibrium density profile. In the following we recapitulate the main aspects of our DFT implementation; for a more in depth explanation we refer to ref.~\cite{glatzel2021reversible}.

For the effective one-dimensional geometry of our system (see Fig.~\ref{Fig:SemiCanonical}), the standard representation of the functional reads \cite{Hansen2013Theory}
\begin{align}\label{eq:GrandFunc}
	\Omega[\{ \rho_i \}] =& \mathcal{F}_\mathrm{id}[\{ \rho_i \}] + \mathcal{F}_\mathrm{exc}[\{ \rho_i \}] \notag \\ &+ \sum_i A \int \rho_i(z) \left[V_i^{\mathrm{ext}}(z) - \mu_i \right] \dd z
\end{align}
and depends on the densities $\rho_i(z)$, the chemical potentials $\mu_i$, and the external potentials $V_i^{\textrm{ext}}(z)$ from eq.~(\ref{eq:Ext_Potential}). Here and in the following, $\mu_i$ can be understood as functions of bulk densities and temperature, $\mu_i(\bar{\rho}_i,T)$, chosen such that the bulk densities attain a predetermined concentration $\bar{\rho}_i=c_i$. Once the chemical potential is determined for a certain bulk density (concentration) and temperature, it is kept constant during minimizing the grand potential functional. Thereby its choice ensures that the bulk concentration of interest is achieved. The translational symmetry along the infinitely extended wall is reflected in its surface area $A$ that enters as a linear scale. The functionals $\mathcal{F}_\mathrm{id}$ and $\mathcal{F}_\mathrm{exc}$ are the ideal and excess free energy functionals, respectively, where the latter contains all contributions from particle interactions beyond the ideal gas.  The ideal free-energy term has the following analytical expression: 
\begin{equation}\label{eq:F_id}
	 \mathcal{F}_\mathrm{id}[\{ \rho_i \}] =\frac{A}{\beta}\sum_i\int \rho_i(z)\left[\ln(\rho_i(z)\Lambda_i^3)-1 \right]\mathrm{d}z,
\end{equation}
where $\Lambda_i$ is the thermal wavelength of the $i$-th particle species. The form of the remaining excess free energy term $\mathcal{F}_\text{exc}$ is not known in general. For our system of charged hard spheres, a common approach is to split the excess term up into two contributions \cite{Haertel2017structure}: \begin{equation}\label{eq:F_ext}
	 \mathcal{F}_\mathrm{exc}[\{ \rho_i \}] = \mathcal{F}_\mathrm{HS}[\{ \rho_i \}] + \mathcal{F}_\mathrm{C}[\{ \rho_i \}].
\end{equation}
In this perturbation around a hard-sphere reference system, the first term $\mathcal{F}_\mathrm{HS}$ covers the hard-sphere repulsion and the second term $\mathcal{F}_\mathrm{C}$ adds Coulomb interactions in a mean-field approximation. 
Note that additional correlations between hard and electrostatic interactions are neglected in this description. As in previous work \cite{glatzel2021reversible}, we use the well-performing White-Bear-mark-II functional 
for the hard-sphere repulsion \cite{HansenGoos2006density,Roth2010fundamental, Oettel2010free} with a correction by Tarazona \cite{Tarazona2000density}. The respective mean-field electrostatic term
\begin{equation}
	\mathcal{F}_\mathrm{C}[\{ \rho_i \}] = \frac{e A}{2} \int q(z)\psi(z) \mathrm{d} z 
\end{equation}
adds the locally evaluated contribution from the charge density $q(z)$ in the electrostatic potential $\Psi(z)$.
Here, the charge density is given by
\begin{equation}
	q(z) = \sum_i z_i \rho_i(z) + \sigma\left[\delta(z)+\delta(L-z)\right],
\end{equation} 
where $\delta(z)$ is the Dirac-$\delta$ distribution. Note that the surface charge density here also enters the functional via the total charge density. The electrostatic potential and the charge density are related via the Poisson equation
\begin{equation}\label{eq:PoissonEqu}
	\epsilon_0\epsilon_{r}\frac{\partial^2 \psi}{\partial z^2} = -e q(z),
\end{equation}
where we use the boundary condition $\lim_{z \to 0}\psi^\prime(z)=-e\sigma/(\epsilon_0\epsilon_r)$.

With equations~(\ref{eq:GrandFunc})-(\ref{eq:PoissonEqu}) the functional derivatives of $\Omega[\{\rho_i\}]$ with respect to the particle densities can be evaluated such that the equilibrium densities read

\begin{align}\label{eq:EulerEq}
    \rho_j^{\textrm{eq}}(z) = \Lambda^{-3}&\exp\left( \left.-\beta\frac{\delta \mathcal{F}_{\textrm{HS}}}{\delta\rho_j(z)}\right|_{\rho_i=\rho_i^{\textrm{eq}}}\right) \nonumber \\
    \times &\exp\left(-\beta z_je\psi(z)-\beta(V^{\textrm{ext}}_j(z)-\mu_j)  \right).
\end{align}
This equation can be solved iteratively, because
$\rho_j^{\textrm{eq}}$ appears on both sides of the equal sign. We apply a simple Picard iteration scheme to solve eq.~(\ref{eq:EulerEq}) self-consistently for the density, as, for instance, explained in the appendix of ref.~\cite{Ng1974Hypernetted}. In the appendix of this work we exemplarily show equilibrium density profiles for symmetric and asymmetric ion models and demonstrate the influence of our HSE model extension on them.

\subsection{Thermodynamic analysis via DFT}
As mentioned earlier, for the following analysis we are interested in the reversible heat and, following eqs.~(\ref{eq:GrandFunc})-(\ref{eq:PoissonEqu}), we evaluate the functional $\Omega[\{\rho_i\}]$ at the numerically obtained equilibrium densities to obtain the grand potential. Thus, we have access to thermodynamic quantities via the respective partial derivatives. For a more in depth discussion than we give in the following, we refer to our previous work~\cite{glatzel2021reversible}.

The semi-canonical process, as introduced before eq.~(\ref{equ:Entropy}), simplifies the calculations in a DFT framework drastically, because it relies on quantities that are naturally accessible via DFT. Consequently, the change in entropy $\Delta S$ can be accessed from the functional $\Omega[\{\rho_i\}]$ via eq.~(\ref{equ:Entropy}), where the derivatives are performed at constant bulk densities instead of chemical potentials such as in the grand-canonical case. To convince oneself that this is in accordance with the experimental setup, one can imagine the whole system to consist of two subsystems, as sketched in Fig.~\ref{Fig:SemiCanonical}. One is comparably small and describes pores of the electrodes and the other one is comparably large and resembles the connected tank of reservoir. By (dis)charging the capacitor or changing the temperature some particles will be transferred from the tank to the pores or vice versa. However, for temperature changes we always assume that both subsystems are still in thermal equilibrium and, thus, there is no thermal driving force moving particles from the bulk of one subsystem to the bulk of the other one. Furthermore, as the tank of reservoir is comparably large, the bulk densities in the tank do not change during (dis)charging and, hence, the bulk densities in the pores do not change either.

By determining $\Delta S$ one also knows the reversible heat $\mathcal{Q}_\mathrm{rev}=T \Delta S$. For convenience, in a semi-canonical charging process one can determine the entropy difference $\Delta S$ alternatively by summing over the following entropic contributions
\begin{equation}\label{Eq:NumEnt}
\Delta S = \frac{1}{T}\left(\Delta\mathcal{F}_\mathrm{id}-\sum_i\mu_i \Delta N_i + \Delta\mathcal{F}_\mathrm{HS}\right),
\end{equation}
where $\Delta\mathcal{F}_\mathrm{id}$ and $\Delta\mathcal{F}_\mathrm{HS}$ are the differences in the ideal free energy functional and the hard sphere functional, specified in eqs.~(\ref{eq:F_id}) and (\ref{eq:F_ext}), and $\Delta N_i$ is the difference of particles of species $i$. The number of particles of species $i$ in the system is given by $N_i=A\int\rho_i(z)\dd z$. All these quantities are either easily accessible or have to be evaluated in a DFT calculation anyway, thus one is able to access the reversible heat without performing the numerical derivative of the grand potential with respect to temperature via eq.~(\ref{equ:Entropy}). However, since eq.~(\ref{Eq:NumEnt}) is empirically motivated and cannot be proven to hold in general it must be numerically verified that both expressions for the entropy, eq.~(\ref{equ:Entropy}) and (\ref{Eq:NumEnt}), are consistent. It was numerically verified in ref.~\cite{glatzel2021reversible} that indeed the entropy can be calculated with eq.~(\ref{Eq:NumEnt}) in the RPM and we affirmed numerically that it holds for asymmetrically sized ions, for the HSE model extension, as well as in cases where dehydration energy is considered explicitly, too.
Note that, for these calculations, we assume the relative electric permittivity to be independent of temperature in order to achieve a consistent microscopic model. Instead of calculating the entropy via eq.~(\ref{equ:Entropy}), one could also obtain it via integrating the difference between internal energy and performed electric work. This consistency would break, if, instead, one would use the empirical relationship for water, $\partial(\varepsilon_r(T)T)/\partial T=0$, as discussed in appendix D of our previous work~\cite{glatzel2021reversible}.

In this context, we also stress that we aim to study the entropic effect of hydration-shell evasion, or, the change of the Stern separation. We do not aim to describe the microscopic structure and buildup of hydration shells by explicitly modeling the ion-dipole and dipole-dipole interactions which are present in a real sodium chloride solution. When we consider explicit steric solvent interactions, we simply fix the volume fraction of the neutral solvent without matching the solvent concentration to that of a realistic solvent. Nevertheless, our hypothesis from our previous work~\cite{glatzel2021reversible} is that the physics of the EDL is mainly determined by the minimal Stern layer separation of ions that we consider in the HSE model.

\section{Results}
\subsection{Reversible heat production in the restricted primitive model}
\label{sec:results-rpm}
\begin{figure}
	\includegraphics[width=\linewidth]{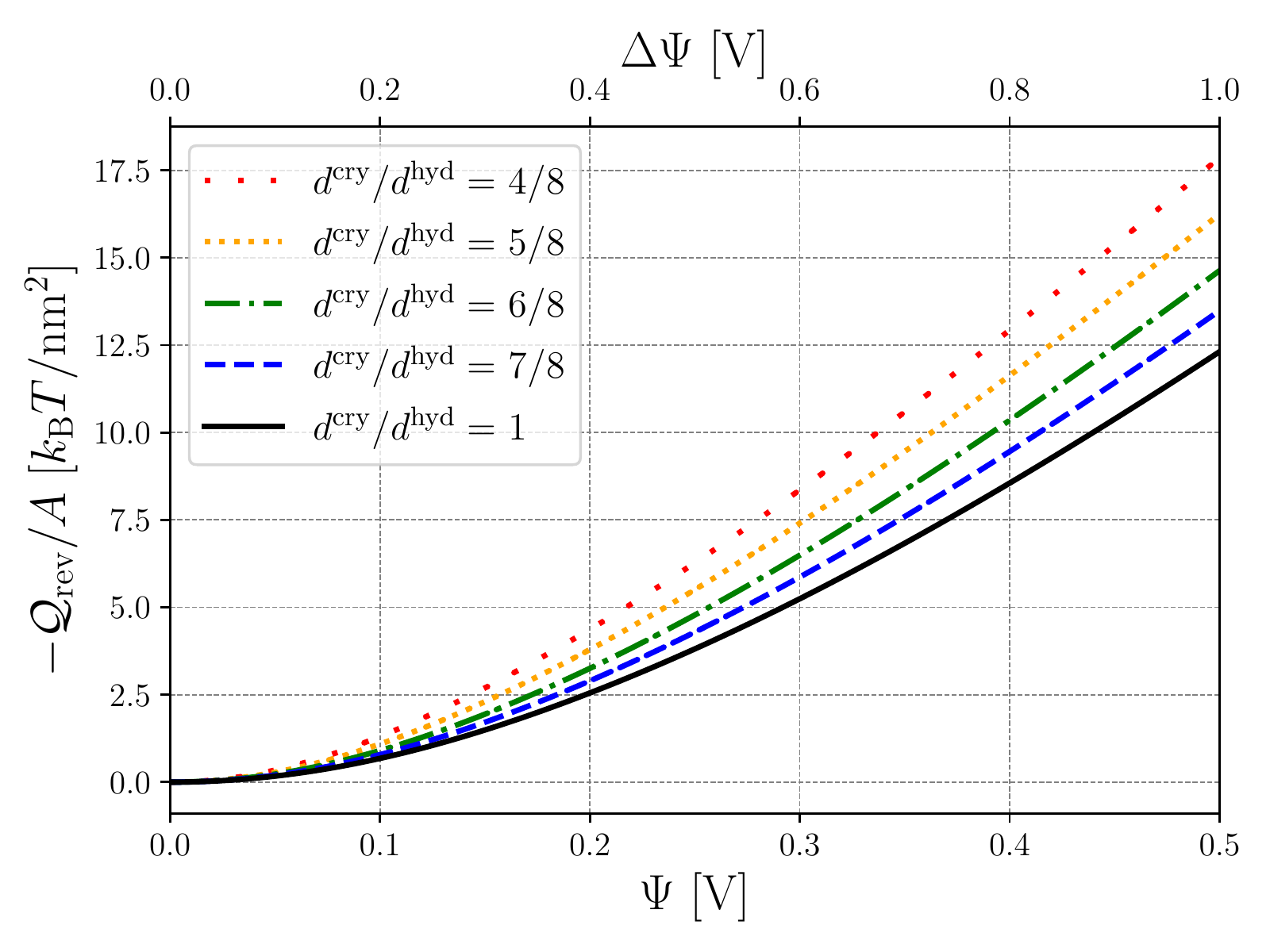}
	\caption{The negative reversible heat in the RPM-HSE against the applied surface potential at one electrode for different ratios of crystallized to hydrated ion diameters, all shown for \SI{1}{\Molar} ion concentration and fixed hydrated diameter $d^\mathrm{hyd}=\SI{0.68}{\nm}$. 
	For convenience, the potential difference $\Delta\Psi$ between both electrodes is shown on a second $x$-axis.}\label{Fig:RPMHeat_w_HSE}
\end{figure}

\begin{figure}
	\includegraphics[width=\linewidth]{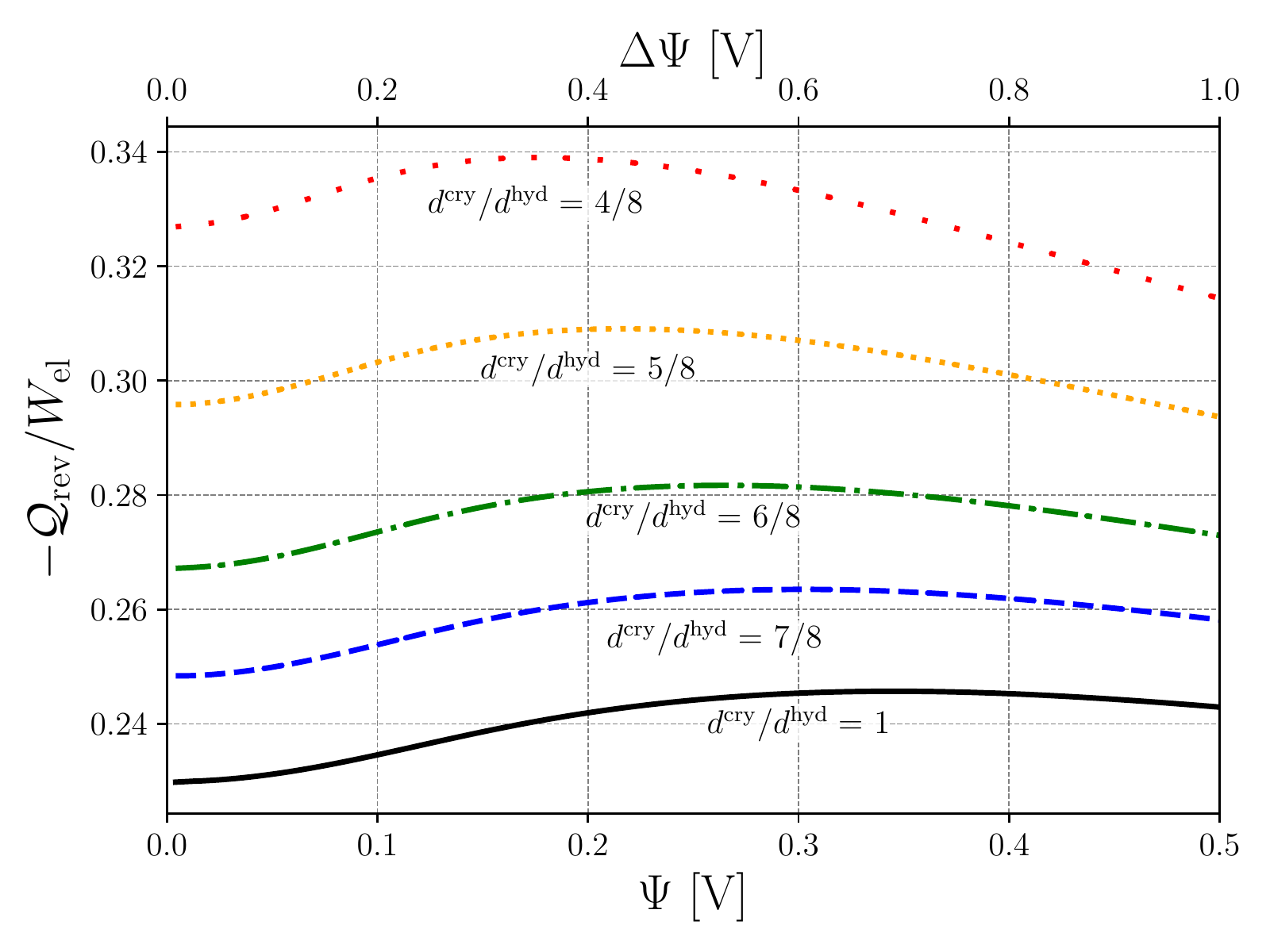}
	\caption{Same as in Fig.~\ref{Fig:RPMHeat_w_HSE}, but here the negative reversible heat per electric work is shown.}\label{Fig:RPMHeatWork_w_HSE}
\end{figure}

\begin{figure}
	\includegraphics[width=\linewidth]{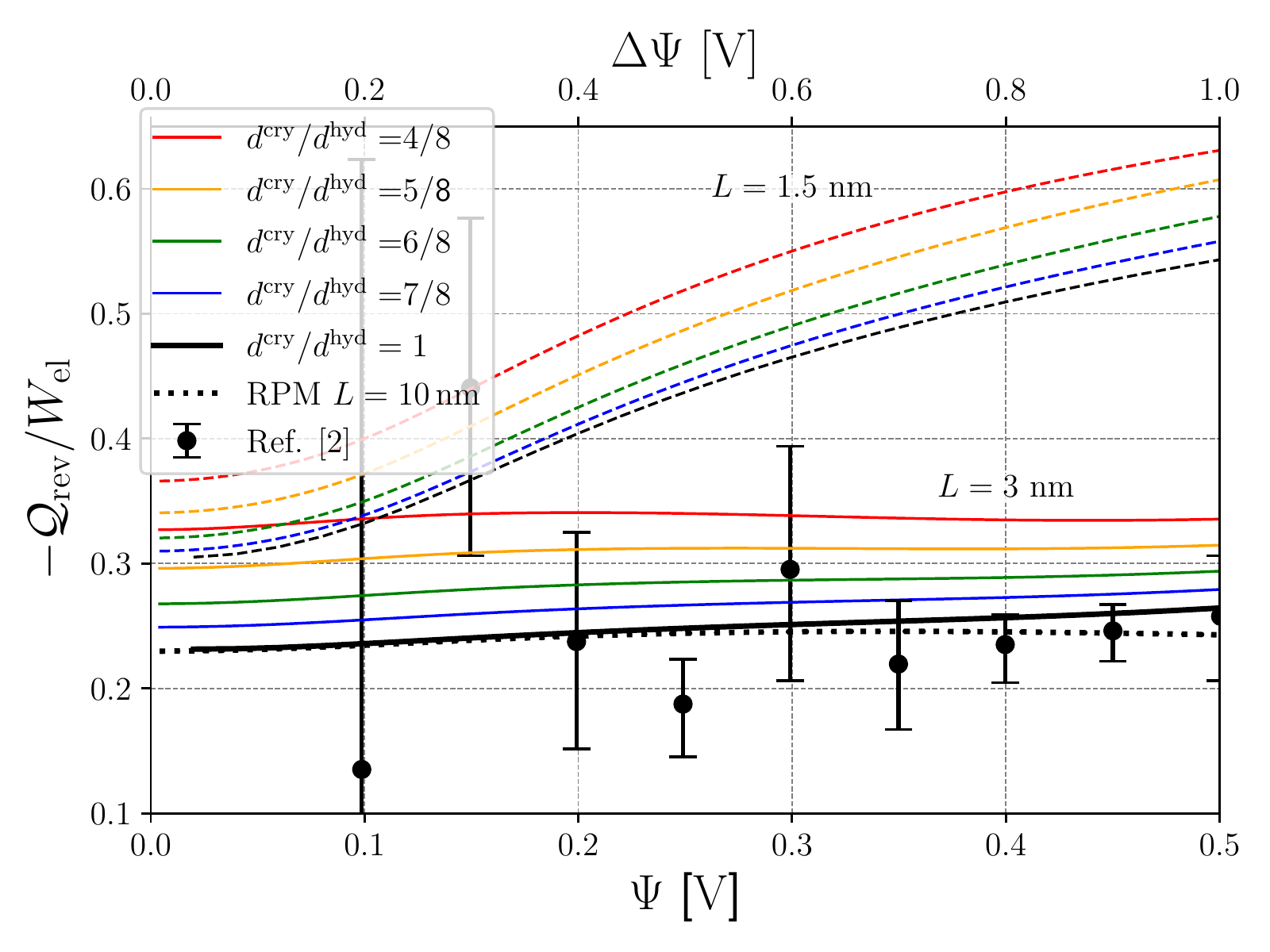}
	\caption{The negative reversible heat per electric work for the applied surface potentials $\Psi$ at one electrode for the RPM-HSE at \SI{1}{\Molar} ion concentration. $\Delta\Psi$ shows the potential difference between both electrodes. Here, EDLs do overlap, because wall separations are small with $L=\SI{3}{\nm}$ (solid lines) and $L=\SI{1.5}{\nm}$ (dashed lines). For comparison the RPM data with  $d^\mathrm{cry}=d^\mathrm{hyd}$ and $L=\SI{10}{\nm}$ (no EDL overlap) from Figs.~\ref{Fig:RPMHeat_w_HSE} and \ref{Fig:RPMHeatWork_w_HSE} are shown as the black dotted line. The symbols show the experimental data from ref.~\cite{janssen2017coulometry}.}\label{Fig:RPMHeat_w_HSE_Lengths}
\end{figure}
We start from the RPM as a reference system to study the effects of our ``hydration-shell evasion'' (HSE) model extension. In order to compare with previous work \cite{glatzel2021reversible}, we choose $d^{\mathrm{hyd}}=d^{\mathrm{cry}}=\SI{0.68}{\nm}$, $c_+ = c_- = \SI{1}{\Molar}$, $L=\SI{10}{nm}$, and $T=\SI{300}{\K}$. Now, we reduce the crystallized diameter $d^{\mathrm{cry}}$ stepwise such that ions effectively have a stepwise increasing hydration shell that can be shed in the vicinity of the wall, as depicted in Fig.~\ref{Fig:HSE} for this RPM-HSE model. In other words, the ions can penetrate the wall with their hydration but not with their core volume (also reflected in the density profiles, exemplarily shown in the appendix of this work).
In Figs.~\ref{Fig:RPMHeat_w_HSE} and \ref{Fig:RPMHeatWork_w_HSE}
we show the results for diameter ratios $d^{\mathrm{cry}}/d^{\mathrm{hyd}}$ being 8/8, 7/8, 6/8, 5/8, and 4/8. Here, the RPM reference with ratio 1 is shown as the black solid line. In Fig.~\ref{Fig:RPMHeat_w_HSE} we show the negative reversible heat per surface area against the electrode surface potential $\Psi$. For equally-sized ions, as in the RPM, the reversible heat and electric work per electrode are the same for positive and negative $\Psi$ such that it is sufficient to show only results for positive $\Psi$.
For a straightforward comparison to experiments and previous work \cite{janssen2017coulometry, glatzel2021reversible} we also added $\Delta\Psi$ on a second $x$-axis. Here, according to the previously mentioned symmetry in the RPM, $\Delta\Psi$ simply equals $2\Psi$. 
As can be seen in Fig.~\ref{Fig:RPMHeat_w_HSE}, the absolute amount of heat increases with decreasing crystallized diameters as smaller cores lead to higher ion concentrations in the EDL and increase the amount of ion sorting during charging. In Fig.~\ref{Fig:RPMHeatWork_w_HSE}, we show the negative reversible heat per electric work against the potential (difference) for the same data as shown in Fig.~\ref{Fig:RPMHeat_w_HSE}.

In ref.~\cite{glatzel2021reversible} it was shown that the heat production within the RPM agrees with the experimental data only for large pore sizes but not for small pore sizes $L<\SI{4}{\nm}$. This observation is unexpected. A surface to volume ratio of the porous electrodes used in the experiment was determined with BET adsorption measurements \cite{janssen2017coulometry}. To obtain the same surface to volume ratio in our model system we have to choose an electrode distance of around $L=\SI{1.6}{\nm}$.
This small wall separation is comparable to the ion size itself and results in overlapping electric double layers. 
The mismatch between the RPM results in ref.~\cite{glatzel2021reversible} and the experimental data was attributed to the presence of hydration shells that are not taken into account in the RPM. Here, we now directly test this hypothesis with the RPM-HSE model and show the results in Fig.~\ref{Fig:RPMHeat_w_HSE_Lengths}. The black dotted curve represents the RPM results as shown in Figs.~\ref{Fig:RPMHeat_w_HSE} and \ref{Fig:RPMHeatWork_w_HSE} for $L=\SI{10}{\nm}$; the solid and dashed curves show results for $L=\SI{3}{\nm}$ and $L=\SI{1.5}{\nm}$, respectively shown for a set of different diameter ratios $d^{\mathrm{cry}}/d^{\mathrm{hyd}}$. Apparently, $-\mathcal{Q}_{\mathrm{rev}}/W_{\mathrm{el}}$ still increases with decreasing crystallized diameters even if EDLs overlap such that 
ion dehydration cannot explain the mismatching data between experiment and theory. However, this result does not proof against ion dehydration in general. The experimentally measured ratio $-\mathcal{Q}_{\mathrm{rev}}/W_{\mathrm{el}}$ from ref.~\cite{janssen2017coulometry} is a combined quantity for both electrodes, where contributions of asymmetric ion sizes and steric ion-solvent interactions cannot be resolved in the data. 
In the following we will demonstrate that particularly asymmetric ion diameters and additional steric ion-solvent interactions take a significant role in the reversible heat production during the EDL buildup. Another explanation in ref.~\cite{glatzel2021reversible} for the mismatching data concerns the pore sizes relevant for the experimental measurement: While the pore sizes in a nanoporous electrode cover several orders of magnitude, it is unknown to which degree each pore contributes to a system observable, such as the reversible heat per electric work. Since all theoretical approaches agree with the experimental data for pore sizes of $L\geq\SI{10}{\nm}$ but not for smaller pore sizes, the contribution of the smaller pores could indeed be of minor significance in this case. For instance, calculating the reversible heat production per surface area from the experimental results \cite{janssen2017coulometry}, one obtains $-\mathcal{Q}_\textrm{rev}/A\approx 0.4$~$k_\textrm{B}T\si{nm^{-2}}$ at $\Psi=\SI{0.5}{V}$ for an ion concentration of \SI{1}{\Molar}. The respective theoretical prediction from Fig.~\ref{Fig:RPMHeat_w_HSE} is around $15$~$k_\textrm{B}T\si{nm^{-2}}$. This discrepancy would be explained by an overestimated surface area determined through the BET measurements in the experiment, for instance, because smaller pores are not accessible for the electrolyte. Another reasonable cause could be clogging of nanopores during the charging process \cite{breitsprecher2017effect, breitsprecher2018charge}. For the rest of this article we limit our discussion 
to the electrode distance $L=\SI{10}{\nm}$ where EDLs do not overlap in order to study effects arising from asymmetric ion models on a single EDL.

\subsection{Reversible heat production in the primitive model}
\label{sec:results-pm}
\begin{figure}
	\includegraphics[width=\linewidth]{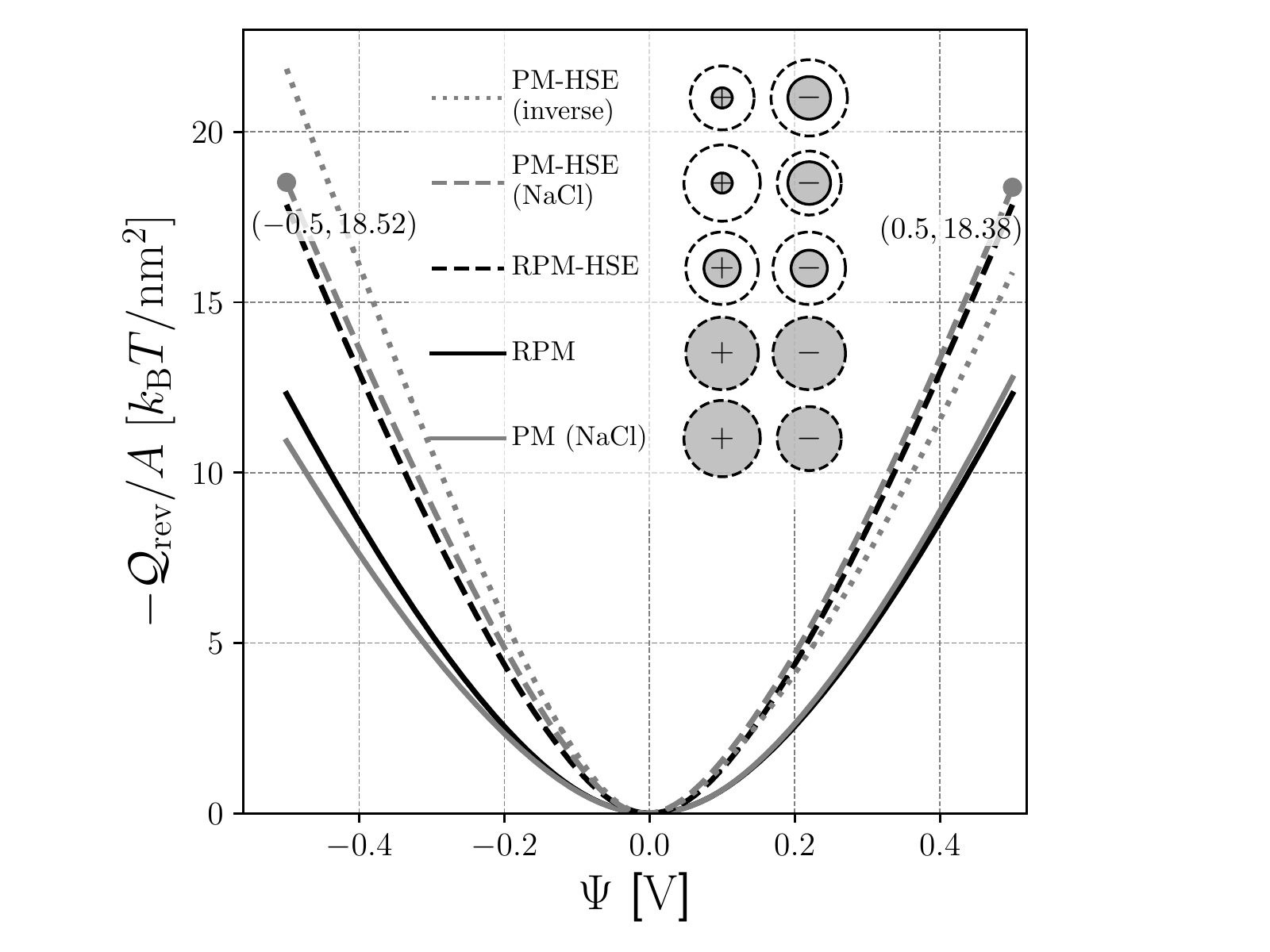}
	\caption{The negative reversible heat against the applied potential at one electrode for different ion models at \SI{1}{\Molar} ion concentration. The continuous curves represent traditional primitive models. The dashed/dotted curves show ion models with HSE extension. The (black continuous) line for RPM and the (black dashed) line for RPM-HSE are the same as in Fig.~\ref{Fig:RPMHeat_w_HSE} for $d^\mathrm{cry}/d^\mathrm{hyd}=1$ and $d^\mathrm{cry}/d^\mathrm{hyd}=1/2$ respectively, and, thus, symmetric for positive and negative potentials. For PM-HSE (NaCl), the numerical values of the heat are shown in two points to demonstrate asymmetry. In the model sketches, the hydrated diameter is shown as the dashed line and, the crystallized diameter is shown as gray circle. }\label{Fig:PMHeat}
\end{figure}

While for equally-sized ions the density profiles and all related physics are symmetric at both electrodes, we have to study each electrode separately for asymmetric ion sizes. Motivated by previous work \cite{janssen2017coulometry}, we use the crystallized and hydrated diameters for aqueous sodium-chloride solutions \cite{nightingale1959phenomenological} as listed in Table~\ref{Tab:IonModels}. In this PM (NaCl) the hydrated sodium ion is larger than the hydrated chloride ion. However, the crystallized diameters are related the other way round, as discussed in sec.~\ref{sec:IonModels}. In the hydration-shell evasion model, now, both diameters are combined such that the larger sodium is able to come closer to the wall than the smaller chloride ion. In order to investigate consequences arising from these different diameters, we additionally study a PM-HSE (inverse) system, where we interchange the hydrated diameters for positive and negative ions in comparison to PM-HSE (NaCl), as depicted in the inset of Fig.~\ref{Fig:PMHeat}.

In Fig.~\ref{Fig:PMHeat}, we show the negative reversible heat $-\mathcal{Q}_\textrm{rev}$ for the different ion models that we discussed. There are two effects which one can conclude from these results. First, $-\mathcal{Q}_\textrm{rev}$ increases with decreasing minimal Stern layer separation, which corresponds to $d^{\textrm{cry}}$. At the same time, larger hydrated diameters reduce the number of ions in the first layer of the EDL and decrease the negative reversible heat (compare PM-HSE (NaCl) and PM-HSE (inverse) in Fig.~\ref{Fig:PMHeat}). Second, asymmetry in the ion diameters is reflected in $\mathcal{Q}_\textrm{rev}$ only marginally.

We mention that the HSE ion models show a weak shift in the open circuit potential $\Psi_{\mathrm{OCP}}\approx\SI{6}{mV}$, which is the surface potential at which the surface charge is zero. 
The minimum of $-\mathcal{Q}_{\mathrm{rev}}$ is shifted slightly to negative potentials.
Both are not visible in Fig.~\ref{Fig:PMHeat}, but will be relevant for the SPM-HSE in Sec.~\ref{sec:results-spm}.

\subsection{Reversible heat production in the solvent primitive model}\label{sec:results-spm}
\begin{figure}
	\includegraphics[width=\linewidth]{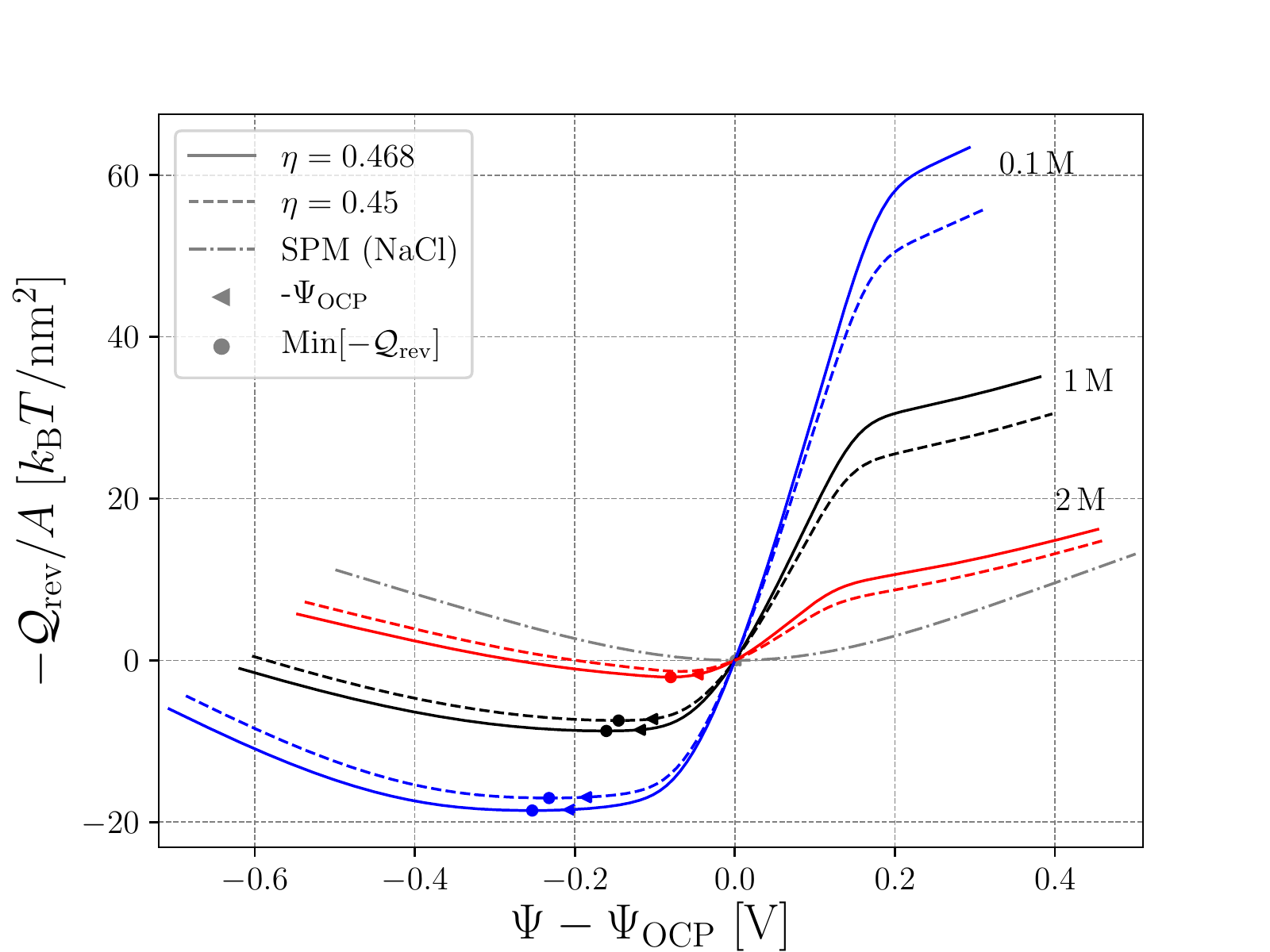}
	\caption{The negative reversible heat against the applied surface potential at one electrode for different ion concentrations for SPM-HSE (NaCl), shifted by $\Psi_\mathrm{OCP}$. The continuous curves are obtained at $\eta=0.468$, and the dashed curves are obtained at $\eta=0.45$. The gray dashed-dotted line shows SPM (NaCl) for \SI{1}{M} ion concentration at $\eta=0.468$. The shift by $\Psi_{\mathrm{OCP}}$ is indicated by a triangle symbol at the corresponding curves. The minimum of $-\mathcal{Q}_{\mathrm{rev}}$ is shown as the circle on the corresponding curves.}\label{Fig:SPM_concs}
\end{figure}
\begin{figure}
	\includegraphics[width=\linewidth]{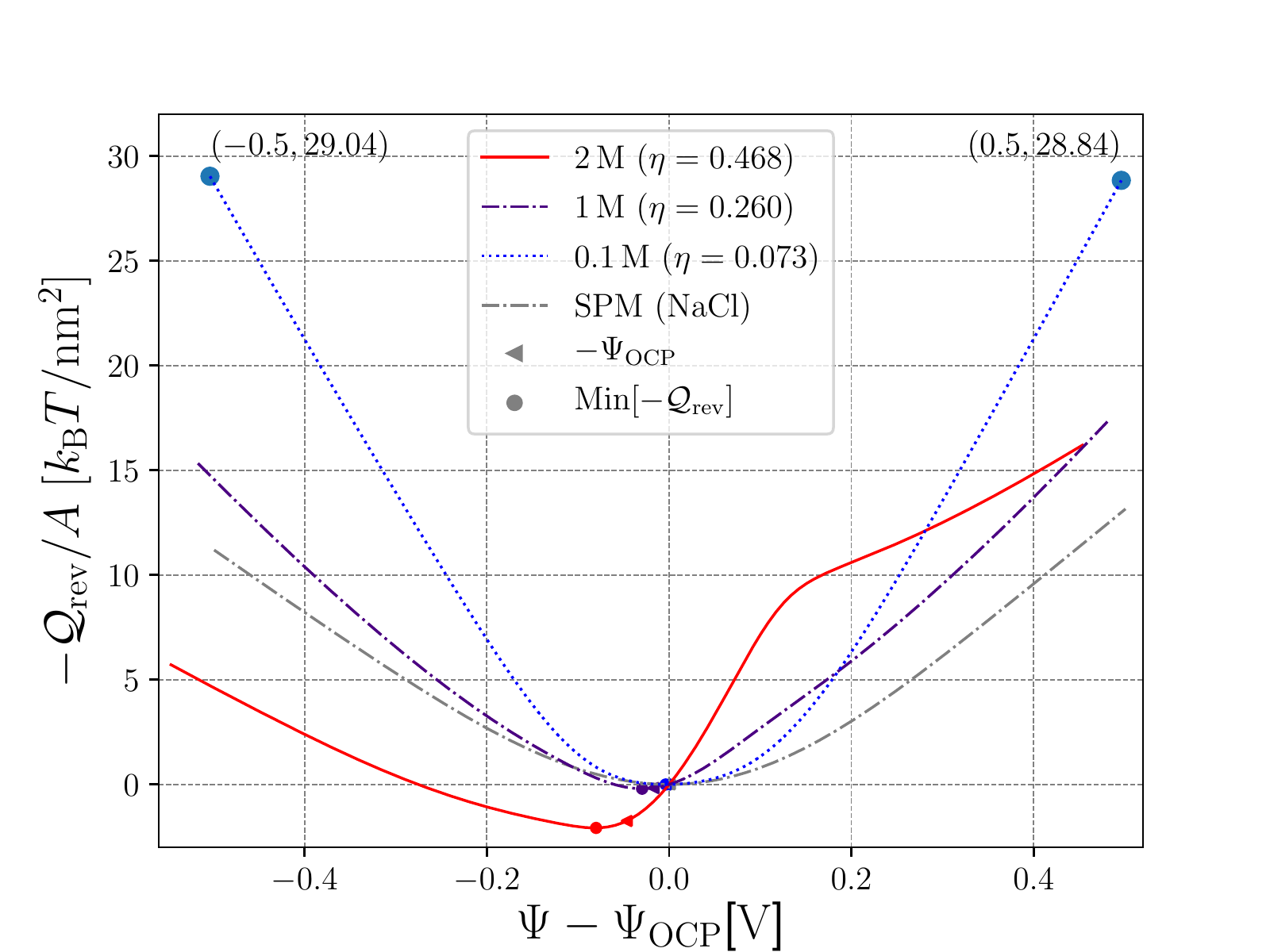}
	\caption{The negative reversible heat against the applied surface potential at one electrode, shifted by $\Psi_\mathrm{OCP}$, for different ion concentrations at fixed solvent concentration $c_\circ=\SI{6}{\Molar}$ for SPM-HSE (NaCl). The red curve $(c_\pm=\SI{2}{M})$ and the gray curve $(c_\pm=\SI{1}{M} \textrm{ at } \eta = 0.468)$  are the same as in Fig.~\ref{Fig:SPM_concs}. The shift by $\Psi_{\mathrm{OCP}}$ is indicated by a triangle symbol at the corresponding curve. The minimum of $-\mathcal{Q}_{\mathrm{rev}}$ is shown as the circle on the corresponding curves.}\label{Fig:SPM_volfrac}
\end{figure}
Real solvents take up volume and sterically interact with the ions. To study its impact, we now add neutral solvent particles to our model such that solvent particles take up volume in this SPM-HSE model. Dielectric contributions by the solvent are still covered implicitly by the dielectric constant. We call the interested reader's attention to the appendix, where we exemplary show respective density profiles. In addition to our present discussion of symmetric and asymmetric heat production, these density profiles demonstrate exemplarily how the related structure of the EDLs changes.

In Fig.~\ref{Fig:SPM_concs} we show the results from the SPM-HSE model for the fixed total volume fraction $\eta$.
As can be seen, the SPM-HSE is sensitive to the value of the total volume fraction. The shape of the reversible heat curves changes drastically as it can be seen from a comparison between the \SI{1}{\Molar} SPM-HSE (NaCl) curves in Fig.~\ref{Fig:SPM_concs} and the curves in Fig.~\ref{Fig:PMHeat}. While in previous discussed ion models the heat curves were ``U-shaped'', they exhibit a characteristic ``S-shape'' in SPM-HSE. At negative potentials the negative reversible heat is much less when a neutral solvent is present, while for positive potentials it can reach much larger values depending on the ion concentration. Additionally, the minima of $-\mathcal{Q}_{\mathrm{rev}}$ are shifted significantly to negative potentials. This effect increases with decreasing ion concentrations. Note that in SPM-HSE the minima do not anymore solely depend on $\Psi_{\mathrm{OCP}}$ such as in the (R)PM-HSE model, even though $\Psi_{\mathrm{OCP}}$ is the largest contribution to its shift. These effects originate from an interplay of the HSE model extension with solvent particles, since neither the plain SPM (NaCl) (grey dash-dotted line in Fig.~\ref{Fig:SPM_concs}) nor the PM-HSE models in Fig.~\ref{Fig:PMHeat} exhibit these traits.  

Since the SPM-HSE strongly depends on the solvent concentration, we now, in Fig.~\ref{Fig:SPM_volfrac}, also take a look at the system's behavior at different ion concentrations for a fixed solvent concentration. We fix the solvent concentration to $c_\circ=\SI{6}{\Molar}$. This solvent concentration is the same 
in Fig.~\ref{Fig:SPM_concs} for the \SI{2}{\Molar} ion concentration curve at $\eta=0.468$. Accordingly, the (red) curve for \SI{2}{\Molar} ion concentration in Fig.~\ref{Fig:SPM_volfrac} is the same as in Fig.~\ref{Fig:SPM_concs} and it has the characteristic SPM-HSE shape. Reducing the ion concentration to \SI{1}{M} (black dashed curve) already turns the reversible heat into ``U-shape'', where the asymmetry is still noticeable. For \SI{0.1}{M}, the asymmetry in the reversible heat is marginal, but flipped in comparison to higher ion concentrations. For low ion concentrations at low volume fractions, the solvent particles approach the ideal gas limit, hence the system resembles PM-HSE without an explicit solvent (compare with gray dashed curve in Fig.~\ref{Fig:PMHeat}).

\subsection{A simple correction for dehydration energy}
\label{sec:Dehydration}
\begin{figure}
	\includegraphics[width=\linewidth]{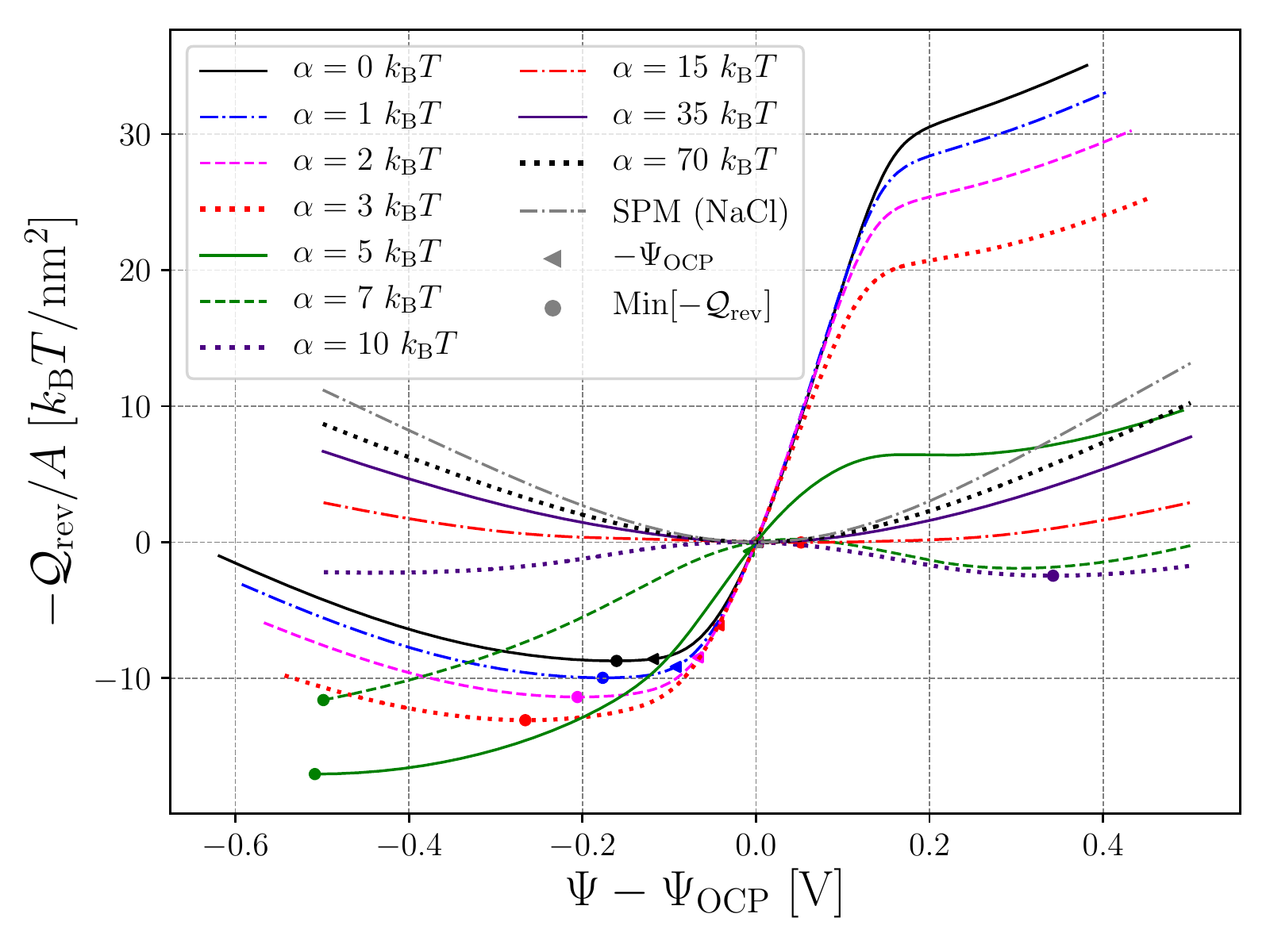}
	\caption{The negative reversible heat against the applied surface potential at one electrode, shifted by $\Psi_\mathrm{OCP}$, at $c=\SI{1}{M}$ for SPM-HSE (NaCl) for different $\mathcal{F}_\mathrm{PEV}$ strengths $\alpha$. The (black) continuous curve is the same as in Fig.~\ref{Fig:SPM_concs} for $c_{\pm}=\SI{1}{M}$ at $\eta=0.468$. The (gray) dashed-dotted line shows SPM (NaCl). The shift by $\Psi_{\mathrm{OCP}}$ is indicated by a triangle symbol at the corresponding curves. The minimum of $-\mathcal{Q}_{\mathrm{rev}}$ is shown as the circle on the corresponding curves.}\label{Fig:SPM_HSEalpha}
\end{figure}
Our HSE model extension does not take into account any form of solvation energy. The ions can just partially permeate the electrodes, whereas in a real physical setting, an ion has to overcome a dehydration free energy to shed its hydration shell and to approach the electrode closer than $d^\mathrm{hyd}/2$, as sketched in Fig.~\ref{Fig:HSE}. In order to estimate possible resulting changes in $-\mathcal{Q}_\textrm{rev}$ due to dehydration energy, we incorporate an additional dehydration free energy functional into the excess functional in eq.~(\ref{eq:F_ext}). We assume that the dehydration free energy is proportional to the
``permeated volume'' that is given by the spherical cap volume
\begin{equation}
	V^\mathrm{per}_i(z) = \frac{\pi h_i(z)^2}{3}\left(\frac{3}{2}d^\mathrm{hyd}_i - h_i(z)\right)
\end{equation}
with the permeation depth 
\begin{equation}
	h_i(z) =
	\begin{cases}
	\tfrac{1}{2}d^\mathrm{hyd}_i - z, &\quad  \tfrac{1}{2}d^\mathrm{cry}_i < z < \tfrac{1}{2}d^\mathrm{hyd}_i \\
	0, &\quad z \geq \tfrac{1}{2}d^\mathrm{hyd}_i . 
	\end{cases}
\end{equation}
We express $V^\mathrm{per}$ in terms of the volume of a solvent particle $V_\circ$ (according to Table~\ref{Tab:IonModels}) and assume that solvating an ion costs an amount $\alpha$ of energy, set in terms of thermal energy $k_\mathrm{B}T$. Consequently, the rescaled permeated volume $V_{i}^\mathrm{per}/V_\circ$ of a particle of species $i$ leads to the additional free energy term 
\begin{equation}
	\mathcal{F}_\mathrm{PEV}[\{ \rho_i \}] = \alpha\frac{A}{V_\circ}\sum_i \int  V^\mathrm{per}_i(z) \rho_i(z)\mathrm{d}z . 
	\label{eq:PEV}
\end{equation}

In Fig.~\ref{Fig:SPM_HSEalpha}, we show the results for the SPM-HSE with incorporated dehydration energy according to eq.~(\ref{eq:PEV}). The limiting case for $\alpha=0$ is the SPM-HSE that has been discussed earlier in Fig.~\ref{Fig:SPM_concs}. For $\alpha>0$ the hydration shell becomes repulsive for the wall and in the limit $\alpha\to\infty$, the SPM (NaCl) is restored.
As one can see in Fig.~\ref{Fig:SPM_HSEalpha}, values of dehydration energy in the order of the thermal energy, i.e. $\alpha < 5$ $k_\textrm{B}T$, have only a minor influence on the reversible heat curves. These curves keep their characteristic shape but are shifted to smaller values of $-\mathcal{Q}_{\textrm{rev}}$. However, in the range $10$~$k_\textrm{B}T \leq \alpha \leq 15$~$k_\textrm{B}T$ a crucial change in the shape of the heat curves occur. In this region $-\mathcal{Q}_{\textrm{rev}}$ transforms from a ``Mexican hat shape'' to a ``U-shape''. Thereby, $-\mathcal{Q}_{\textrm{rev}}$ is comparably flat and minimized for both positive and negative surface potentials. Notably, in this region, the reversible heat increases at both electrodes by charging and, thereby, the entropy difference increases as well.

To estimate the order of magnitude of $\alpha$ for real ions we use the experimentally attained Gibbs free energies of solvation from ref.~\cite{Marcus1991thermodynamics} for sodium and chloride which are reported to be around $-\SI{350}{\kilo\J\mole^{-1}}\approx -140 k_{\textrm{B}}T$ per  ion. We relate this energy to the full hydration-shell volume of one ion according to the used NaCl diameters in our model (Table~\ref{Tab:IonModels}) and find a value of $\alpha \approx 14$ $k_\textrm{B}T$ for a volume $V_{\circ}$. According to Fig.~\ref{Fig:SPM_HSEalpha} this value lies exactly in the region where the reversible heat production is small. Consequently, further studies of dehydration seem to be worthwhile where the hydration process is modeled in more detail.

\subsection{Combining both electrodes}
\begin{figure*}
	\includegraphics[width=0.65\linewidth]{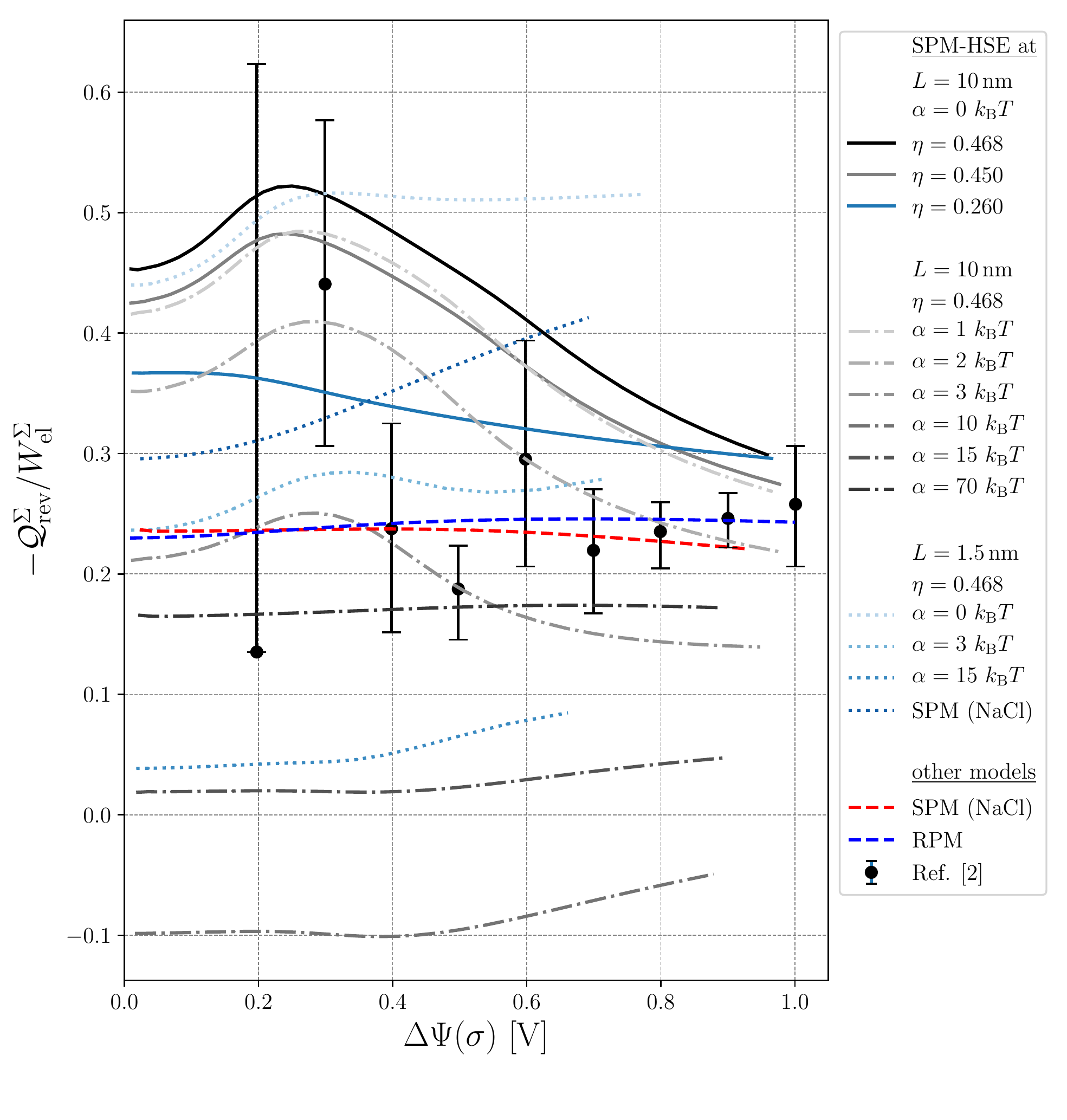}
	\caption{The combined negative reversible heat per combined electric work against the potential difference $\Delta\Psi(\sigma)$ at the electrodes for \SI{1}{\Molar} ion concentration.
	The potential difference $\Delta\Psi(\sigma)$ results from the potentials $\Psi(\pm\sigma)$ at the two electrodes with surface charge densities $\pm\sigma$. 
	Different line styles indicate different groups of parameters: solid lines represent SPM-HSE without energy correction at different total volume fractions $\eta$, as shown for individual electrodes in Figs.~\ref{Fig:SPM_concs} and \ref{Fig:SPM_volfrac}; dash-dotted lines represent SPM-HSE at $\eta=0.468$ with energy correction (thus $\alpha>0$), as shown for individual electrodes in Fig.~\ref{Fig:SPM_HSEalpha}; dotted lines represent selected SPM-HSE results with energy correction but with a smaller pore width $L=\SI{1.5}{\nm}$; dashed lines and symbols represent SPM (NaCl), as shown for individual electrodes in Fig.~\ref{Fig:SPM_concs} and \ref{Fig:SPM_volfrac}, RPM, as shown in Figs.~\ref{Fig:RPMHeatWork_w_HSE} and \ref{Fig:RPMHeat_w_HSE_Lengths}, and experimental results from ref.~\cite{janssen2017coulometry} as shown in Fig.~\ref{Fig:RPMHeat_w_HSE_Lengths}. 
	Note that the limits of vanishing and infinite $\alpha$ in the SPM-HSE with energy correction for $L=\SI{10}{\nm}$ are given by the back solid line (SPM-HSE with $\alpha=0$ and $\eta=0.468$) and the red dashed line (SPM (NaCl) without HSE). The same holds for $L=\SI{1.5}{\nm}$ where the limits are given by the lightest blue dotted line and the darkest blue dotted line, respectively.
	The errorbar for the experimental data at $\Delta\Psi(\sigma)=\SI{0.2}{V}$ was not shown completely in ref.~\cite{janssen2017coulometry} such that we skip the negative errorbar here.} \label{Fig:SPM_heat_work}
\end{figure*}
In previous experimental and theoretical work, the reversible heat has been discussed for both electrodes at once \cite{janssen2017coulometry,glatzel2021reversible}. Furthermore, it has been shown that the symmetric RPM can describe the experimental results within the expected error margin (compare, for instance, Fig.~\ref{Fig:RPMHeat_w_HSE_Lengths}). However, in this work we have shown that large differences in $-\mathcal{Q}_\textrm{rev}$ for the individual electrodes can occur depending on the model ingredients. 
These additional ingredients should make the model more realistic, e.g., by incorporating ionic sizes from scattering measurements. Here, one increases the model complexity and, at the same time, one also assumes to increase the model's predictive capabilities. Since the latter is not guaranteed, each additional ingredient should be tested.

To compare our results of the different ion models to the available experimental results \cite{janssen2017coulometry}, we now have to combine both electrodes and add up heat and work contributions from oppositely-charged electrodes. In this sense, we summarize our previously discussed results in Fig.~\ref{Fig:SPM_heat_work}, but, now, added for both electrodes. For this purpose, we evaluate the reversible heat $\mathcal{Q}_\textrm{rev}$ and electric work $W_\textrm{el}$ as functions of the surface charge $\sigma(\Psi)$ and sum up their values for opposing surface charges, more specifically $\mathcal{Q}^{\Sigma}_\textrm{rev}(\sigma)= \mathcal{Q}_\textrm{rev}(\sigma) + \mathcal{Q}_\textrm{rev}(-\sigma)$ and $W^{\Sigma}_\textrm{el}(\sigma)= W_\textrm{el}(\sigma) + W_\textrm{el}(-\sigma)$. Consequently, $\Delta\Psi(\sigma)=|\Psi(\sigma)-\Psi(-\sigma)|$. Note that the previously presented data for positive and negative potentials do not yield access to the full range of $\Delta\Psi(\sigma)$ on the $x$-axis in Fig.~\ref{Fig:SPM_heat_work} due to the asymmetry in $\Psi(\sigma)$.

As one can see in Fig.~\ref{Fig:SPM_heat_work}, the SPM and SPM-HSE model predictions fit to the available data considering the experimental accuracy. High solvent particle volume fractions shift $-\mathcal{Q}^{\Sigma}_\textrm{rev}/W^{\Sigma}_\textrm{el}$ to larger values, worsening the model's prediction (continuous lines). This shift can be suppressed by including a dehydration energy cost with $\alpha > 0$ (dashed-dotted lines). Depending on $\alpha$ one gets largely different outcomes. Especially in the range of 5 $k_\textrm{B}T < \alpha < 15$ $k_\textrm{B}T$ our results systematically lie below the experimental data and they can even reach negative values of $-\mathcal{Q}^{\Sigma}_\textrm{rev}/W^{\Sigma}_\textrm{el}$. This is the region where we find endothermic charging as discussed along Fig.~\ref{Fig:SPM_HSEalpha}. Interestingly, from a comparison with experimentally available solvation energy data \cite{Marcus1991thermodynamics}, we expect $\alpha$ to lie approximately in this region.

We stress that all discussed theoretical predictions are obtained at an electrode separation of $L=\SI{10}{\nm}$. Note that, in order to resemble the same surface to volume ratio as in the experiment we rather should use $L \approx \SI{1.6}{\nm}$, as discussed in Sec.~\ref{sec:results-rpm}. In Fig.~\ref{Fig:RPMHeat_w_HSE_Lengths}, we have seen for the RPM(-HSE) that a smaller electrode separation shifts $-\mathcal{Q}_\textrm{rev}/W_\textrm{el}$ to higher values, particularly for larger potential differences. Now, in Fig.~\ref{Fig:SPM_heat_work}, we show predictions for $L=\SI{1.5}{\nm}$ in addition to $L=\SI{10}{\nm}$ for selected values of $\alpha$ in the SPM-HSE with hydration cost correction (dotted lines).
Small electrode separations and thus overlapping EDLs result in a much more complex behaviour in the SPM-HSE in comparison to RPM-HSE. The change of  $-\mathcal{Q}_\textrm{rev}/W_\textrm{el}$ in comparison to the $L=\SI{10}{\nm}$ system can not be generalized to be a shift to higher values anymore. However, the statement that $-\mathcal{Q}_\textrm{rev}/W_\textrm{el}$ increases with larger potential differences still applies.

In conclusion, it is not possible to elaborate which of our model ingredients improve the model's predictive capabilities when we sum up the reversible heat over both electrodes and compare with the corresponding available experimental results, because resulting deviations between different models are not significant in this representation. However, we found significant differences in the reversible heat production for our models at individual electrodes. These differences should be experimentally attainable if temperatures are also measured in experiments at individual electrodes.

\section{Summary and conclusion}
In this work we discussed the impact of asymmetrically-sized ions, hydration-shell evasion, and related solvation energies on the reversible heat production in EDLs established by aqueous sodium chloride solutions. For this study, we applied the (solvent) primitive model of electrolytes in the framework of classical DFT and added a simple HSE model extension. In contrast to previous work \cite{glatzel2021reversible}, we studied the reversible heat production for individual electrodes and found huge quantitative and qualitative deviations between our results when we took different model ingredients into account. We demonstrated the importance of treating both electrodes individually over summing up their contributions. In the following we summarize four important results from our work. 

First, previous work \cite{glatzel2021reversible} found that large pores fit the experimental results from ref.~\cite{janssen2017coulometry} better than small pores, while the experiments report a rather small average pore size. We found that hydration-shell evasion (HSE) cannot explain this discrepancy. Instead, the production of negative reversible heat increases when RPM-HSE models are used and, thus, even worsen the theoretical predictions. This is not unexpected, because the assumption of hydration shells decreases the minimal Stern layer separation which has been shown to mainly determine the physics of the system, including the production of negative reversible heat; the latter increases for decreasing ion sizes and, thus, minimal Stern layer separation. Nevertheless, our work confirms the hypothesis that mainly the Stern layer sets physical properties of EDLs. This is supported by a recent study of the differential capacitance of EDLs that finds the same conclusion \cite{Cats2021differential}. Finally, we repeat the previously proposed idea that smaller pores might be less significant for heat production, because they are less accessible for mobile charges in relevant times \cite{breitsprecher2017effect, breitsprecher2018charge}. Hence, if our hypothesis is correct, an experiment with a porous electrode but without the smaller pores, should lead to the same results as in ref.~\cite{janssen2017coulometry}. An ideal test would, of course, be the direct measurement with a single EDL at a flat electrode~\cite{Lindner2019Entropy}, but, due to its small surface compared to the porous electrode, we expect the measurement to be challenging.

Second, we found that differences in ionic diameters across different species have only minor impact on the asymmetry of reversible heat at positive and negative potentials. This fact does not change if we extend the (R)PM by the HSE extension. However, this statement changes completely if we additionally consider volume filling solvent particles in our SPM-HSE. Now, asymmetries in the hydrated and core diameters of ions are strongly reflected in the reversible heat production. We even identified situations, where charging not decreased but increased the reversible heat and, thus, the entropy in the system. Furthermore, the volume-taking solvent results in a significant shift of the open circuit potential $\Psi_\textrm{OCP}$ to positive potentials. All the observed results are sensitive to the total volume fraction of the system. 

Third, we introduced a simple correction with respect to the cost of energy when hydration shells are partly shed in the vicinity of the electrode wall. The reversible heat is sensitive to this correction in particular in a region where our assumed solvation energy meets via simple arguments the experimentally measured total dehydration energy of ions. In this region of interest, the reversible heat not only switches sign, but also takes small values in comparison to the heat calculated using very small or very large correction terms. Thus, we conclude that future studies of hydration-shell evasion could be interesting where the (de)hydration process is modeled in microscopic detail. 

Fourth and finally, we studied the reversible heat production at different electrodes for models with different theoretical ingredients: ionic size asymmetry due to scattering measurement, hydration-shell evasion, explicit volume-taking solvent, and an energetic correction for HSE. Our results differ significantly, not only in quantity but also in quality. However, if we sum up contributions for oppositely-charged electrodes we found that most results fit the available experimental measurements within the reported precision. Thus, future measurements are needed for the reversible heat production at individual electrodes to confirm or reject our results for certain model ingredients. In consequence, our work contributes a guide to experimentally determine the primary model ingredients for the theoretical description of electric double layers. By this, our work in combination with future experiments will lead to a deeper understanding of the physics of EDLs.
 
\section*{Acknowledgement}
We thank B. Ern\'e and D. Gillespie for helpful discussions. 
P.P. and A.H. acknowledge funding from the German Research Foundation (DFG) through Project No. 406121234; F.G. acknowledges funding from the DFG through Project No. 430195928.

\section*{Data Availability}
The data that support the findings of this study are available
from the authors upon reasonable request.

\appendix
\section{Density profiles}
\begin{figure*}[t!	]
	\includegraphics[width=0.8\linewidth]{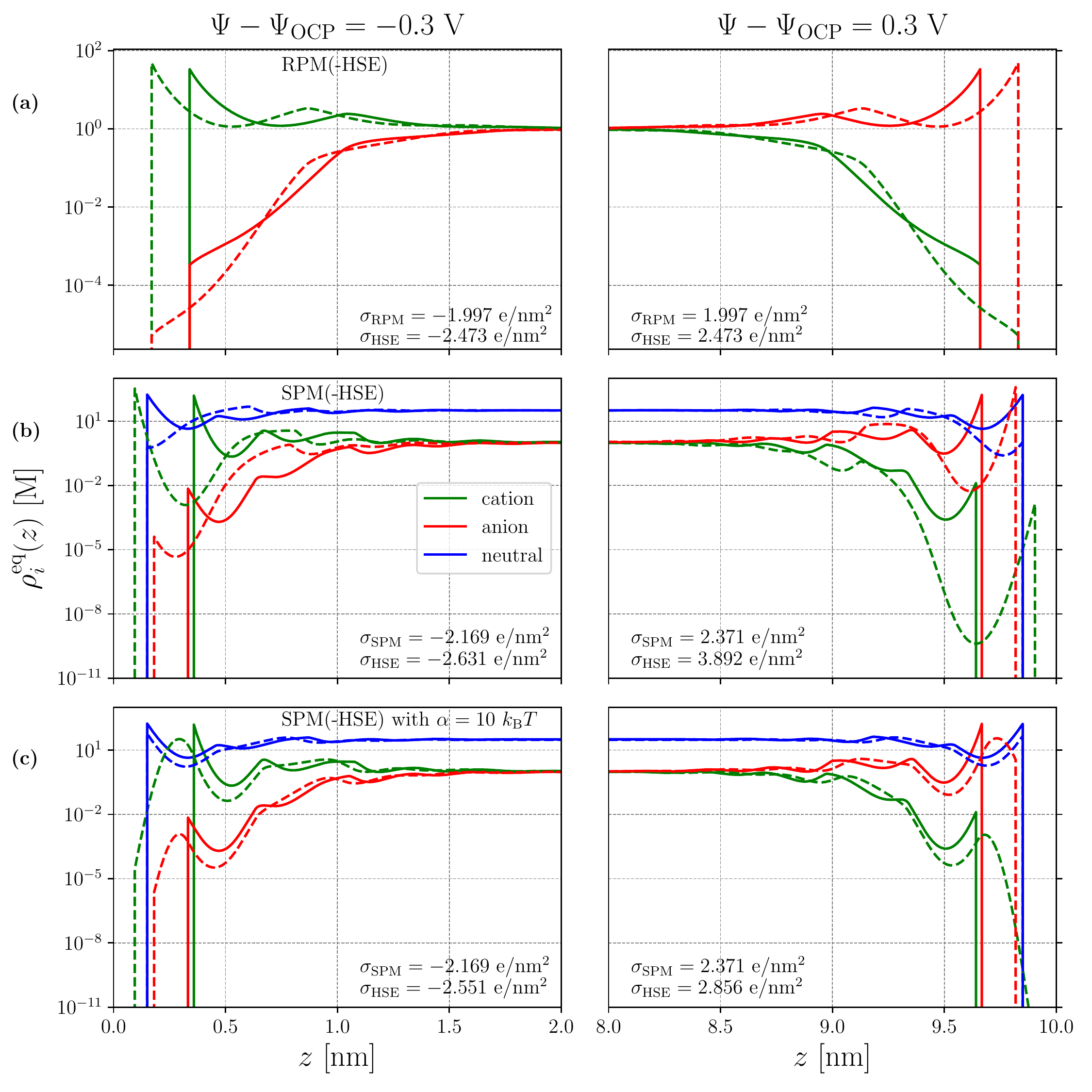}
	\caption{Equilibrium density profiles for different ion models at $\SI{1}{M}$ ion concentration and at an electrode distance of $L=\SI{10}{\nm}$. The continuous lines show  RPM/SPM and the dashed lines represent ion models with HSE extension. The left column shows density profiles for a surface potential of $\Psi-\Psi_{\textrm{OCP}}=\SI{-0.3}{V}$ and the right column for $\Psi-\Psi_{\textrm{OCP}}=\SI{0.3}{V}$. The corresponding surface charges are noted in the bottom corners. \textbf{(a)} RPM with $d=\SI{0.68}{\nm}$ and RPM-HSE with $d^{\textrm{cry}}/d^{\textrm{hyd}}=0.5$. \textbf{(b)} SPM (NaCl) and SPM-HSE at $\eta=0.468$; see Table~\ref{Tab:IonModels} for diameter values. \textbf{(c)} Same as in (b) but SPM-HSE considers an additional dehydration energy cost with $\alpha=10$~$k_\textrm{B}T$. } \label{Fig:DensityProfiles}
\end{figure*}
In Fig.~\ref{Fig:DensityProfiles} representative density profiles for symmetric and asymmetric ion models are shown. Note that, the $y$-axis is logarithmic and spans several orders of magnitude. As one can see for the RPM in Fig.~\ref{Fig:DensityProfiles}(a) the HSE model extension makes the ions approach the electrodes more closely, as constructed. The closer distance also leads to a larger contact value of the counter ions due to electrostatic forces with the electrodes. However, for asymmetric ion models the changes induced by the HSE model extension are not as straightforward as different effects compete with one another as discussed in \ref{sec:results-pm}. In the case of SPM, shown in Fig.~\ref{Fig:DensityProfiles}(b), the contact value of the neutral particles decreases significantly at the negatively-charged electrode when HSE is introduced. This may be attributed to the fact that $d^{\textrm{cry}}_{+} = \SI{0.19}{\nm} < d_\circ = \SI{0.3}{nm}$, such that cations are favored in the proximity of the electrode entropically and electrostatically. However, the contact value of the neutral particles decreases significantly for the positively-charged electrode, too, where this explanation does not hold anymore. This points to the fact, that the observed thermal behaviour of the system is not easily explained by a simple structural argument, but emerges from an interplay between entropic and energetic contributions. Additionally, the asymmetry in ion diameters leads to different density profiles at oppositely-charged electrodes, hence the electrodes must be analyzed separately. In Fig.~\ref{Fig:DensityProfiles}(c) we additionally consider a dehydration energy cost as discussed in \ref{sec:Dehydration}. This makes the density profiles of SPM-HSE converge with increasing $\alpha$ to the ones of SPM, as constructed, and they become more symmetric between the negatively and positively-charged electrodes. The more symmetric density profiles between the electrodes also result in a more symmetric reversible heat curve, as can be seen in Fig.~\ref{Fig:SPM_HSEalpha} for values of $\alpha > 10$ $k_{\textrm{B}}T$.


\newpage
\begin{thebibliography}{54}
	
	\providecommand{\url}[1]{\texttt{#1}}
	\expandafter\ifx\csname urlstyle\endcsname\relax
	\providecommand{\doi}[1]{doi: #1}\else
	\providecommand{\doi}{doi: \begingroup \urlstyle{rm}\Url}\fi
	
	\bibitem[1]{schiffer2006heat}
	\textsc{Schiffer}, Julia ; \textsc{Linzen}, Dirk  ; \textsc{Sauer}, Dirk~U.:
	\newblock Heat generation in double layer capacitors.
	\newblock {In: }\emph{Journal of Power Sources} 160 (2006), September, Nr. 1,
	765--772.
	\newblock \url{http://dx.doi.org/10.1016/j.jpowsour.2005.12.070}. --
	\newblock DOI 10.1016/j.jpowsour.2005.12.070
	
	\bibitem[2]{janssen2017coulometry}
	\textsc{Janssen}, Mathijs ; \textsc{Griffioen}, Elian ; \textsc{Biesheuvel},
	P.~M. ; \textsc{Roij}, Ren{\'{e}} van  ; \textsc{Ern{\'{e}}}, Ben:
	\newblock Coulometry and Calorimetry of Electric Double Layer Formation in
	Porous Electrodes.
	\newblock {In: }\emph{Physical Review Letters} 119 (2017), Oktober, Nr. 16,
	166002.
	\newblock \url{http://dx.doi.org/10.1103/physrevlett.119.166002}. --
	\newblock DOI 10.1103/physrevlett.119.166002
	
	\bibitem[3]{Theodoor1990role}
	\textsc{Theodoor}, J. ; \textsc{Overbeek}, G.:
	\newblock The role of energy and entropy in the electrical double layer.
	\newblock {In: }\emph{Colloids and Surfaces} 51 (1990), Januar, 61--75.
	\newblock \url{http://dx.doi.org/10.1016/0166-6622(90)80132-n}. --
	\newblock DOI 10.1016/0166--6622(90)80132--n
	
	\bibitem[4]{Biesheuvel2007counterion}
	\textsc{Biesheuvel}, P.~M. ; \textsc{Soestbergen}, M. van:
	\newblock Counterion volume effects in mixed electrical double layers.
	\newblock {In: }\emph{Journal of Colloid and Interface Science} 316 (2007),
	Dezember, Nr. 2, 490--499.
	\newblock \url{http://dx.doi.org/10.1016/j.jcis.2007.08.006}. --
	\newblock DOI 10.1016/j.jcis.2007.08.006
	
	\bibitem[5]{dEntremont2014first}
	\textsc{d{\textquotesingle}Entremont}, Anna ; \textsc{Pilon}, Laurent:
	\newblock First-principles thermal modeling of electric double layer capacitors
	under constant-current cycling.
	\newblock {In: }\emph{Journal of Power Sources} 246 (2014), Januar, 887--898.
	\newblock \url{http://dx.doi.org/10.1016/j.jpowsour.2013.08.024}. --
	\newblock DOI 10.1016/j.jpowsour.2013.08.024
	
	\bibitem[6]{dEntremont2015thermal}
	\textsc{d{\textquotesingle}Entremont}, Anna~L. ; \textsc{Pilon}, Laurent:
	\newblock Thermal effects of asymmetric electrolytes in electric double layer
	capacitors.
	\newblock {In: }\emph{Journal of Power Sources} 273 (2015), Januar, 196--209.
	\newblock \url{http://dx.doi.org/10.1016/j.jpowsour.2014.09.080}. --
	\newblock DOI 10.1016/j.jpowsour.2014.09.080
	
	\bibitem[7]{janssen2017reversible}
	\textsc{Janssen}, Mathijs ; \textsc{Roij}, Ren{\'{e}} van:
	\newblock Reversible Heating in Electric Double Layer Capacitors.
	\newblock {In: }\emph{Physical Review Letters} 118 (2017), M{\^^b a}rz, Nr. 9,
	096001.
	\newblock \url{http://dx.doi.org/10.1103/physrevlett.118.096001}. --
	\newblock DOI 10.1103/physrevlett.118.096001
	
	\bibitem[8]{cruz2018electrical}
	\textsc{Cruz}, Carolina ; \textsc{Ciach}, Alina ; \textsc{Lomba}, Enrique  ;
	\textsc{Kondrat}, Svyatoslav:
	\newblock Electrical Double Layers Close to Ionic Liquid{\textendash}Solvent
	Demixing.
	\newblock {In: }\emph{The Journal of Physical Chemistry C} 123 (2018),
	Dezember, Nr. 3, 1596--1601.
	\newblock \url{http://dx.doi.org/10.1021/acs.jpcc.8b09772}. --
	\newblock DOI 10.1021/acs.jpcc.8b09772
	
	\bibitem[9]{alizadeh2020temperature}
	\textsc{Alizadeh}, Amer ; \textsc{Wang}, Moran:
	\newblock Temperature effects on electrical double layer at solid-aqueous
	solution interface.
	\newblock {In: }\emph{{ELECTROPHORESIS}} 41 (2020), Mai, Nr. 12, 1067--1072.
	\newblock \url{http://dx.doi.org/10.1002/elps.201900354}. --
	\newblock DOI 10.1002/elps.201900354
	
	\bibitem[10]{glatzel2021reversible}
	\textsc{Glatzel}, Fabian ; \textsc{Janssen}, Mathijs  ; \textsc{H\"{a}rtel},
	Andreas:
	\newblock Reversible heat production during electric double layer buildup
	depends sensitively on the electrolyte and its reservoir.
	\newblock {In: }\emph{The Journal of Chemical Physics} 154 (2021), Februar, Nr.
	6, 064901.
	\newblock \url{http://dx.doi.org/10.1063/5.0037218}. --
	\newblock DOI 10.1063/5.0037218
	
	\bibitem[11]{Beidaghi2014capacitive}
	\textsc{Beidaghi}, Majid ; \textsc{Gogotsi}, Yury:
	\newblock Capacitive energy storage in micro-scale devices: recent advances in
	design and fabrication of micro-supercapacitors.
	\newblock {In: }\emph{Energy {\&} Environmental Science} 7 (2014), Nr. 3, 867.
	\newblock \url{http://dx.doi.org/10.1039/c3ee43526a}. --
	\newblock DOI 10.1039/c3ee43526a
	
	\bibitem[12]{Suss2015water}
	\textsc{Suss}, M.~E. ; \textsc{Porada}, S. ; \textsc{Sun}, X. ;
	\textsc{Biesheuvel}, P.~M. ; \textsc{Yoon}, J.  ; \textsc{Presser}, V.:
	\newblock Water desalination via capacitive deionization: what is it and what
	can we expect from it?
	\newblock {In: }\emph{Energy {\&} Environmental Science} 8 (2015), Nr. 8,
	2296--2319.
	\newblock \url{http://dx.doi.org/10.1039/c5ee00519a}. --
	\newblock DOI 10.1039/c5ee00519a
	
	\bibitem[13]{Kim2015enhanced}
	\textsc{Kim}, T. ; \textsc{Dykstra}, J.~E. ; \textsc{Porada}, S. ;
	\textsc{Wal}, A. van~d. ; \textsc{Yoon}, J.  ; \textsc{Biesheuvel}, P.~M.:
	\newblock Enhanced charge efficiency and reduced energy use in capacitive
	deionization by increasing the discharge voltage.
	\newblock {In: }\emph{Journal of Colloid and Interface Science} 446 (2015),
	Mai, 317--326.
	\newblock \url{http://dx.doi.org/10.1016/j.jcis.2014.08.041}. --
	\newblock DOI 10.1016/j.jcis.2014.08.041
	
	\bibitem[14]{Brogioli2009extracting}
	\textsc{Brogioli}, Doriano:
	\newblock Extracting Renewable Energy from a Salinity Difference Using a
	Capacitor.
	\newblock {In: }\emph{Physical Review Letters} 103 (2009), Juli, Nr. 5, 058501.
	\newblock \url{http://dx.doi.org/10.1103/physrevlett.103.058501}. --
	\newblock DOI 10.1103/physrevlett.103.058501
	
	\bibitem[15]{Jia2014blue}
	\textsc{Jia}, Zhijun ; \textsc{Wang}, Baoguo ; \textsc{Song}, Shiqiang  ;
	\textsc{Fan}, Yongsheng:
	\newblock Blue energy: Current technologies for sustainable power generation
	from water salinity gradient.
	\newblock {In: }\emph{Renewable and Sustainable Energy Reviews} 31 (2014),
	M{\^^b a}rz, 91--100.
	\newblock \url{http://dx.doi.org/10.1016/j.rser.2013.11.049}. --
	\newblock DOI 10.1016/j.rser.2013.11.049
	
	\bibitem[16]{Janssen2014boosting}
	\textsc{Janssen}, Mathijs ; \textsc{H\"{a}rtel}, Andreas  ; \textsc{Roij},
	Ren{\'{e}} van:
	\newblock Boosting Capacitive Blue-Energy and Desalination Devices with Waste
	Heat.
	\newblock {In: }\emph{Physical Review Letters} 113 (2014), Dezember, Nr. 26,
	268501.
	\newblock \url{http://dx.doi.org/10.1103/physrevlett.113.268501}. --
	\newblock DOI 10.1103/physrevlett.113.268501
	
	\bibitem[17]{Haertel2015heat}
	\textsc{H\"{a}rtel}, Andreas ; \textsc{Janssen}, Mathijs ; \textsc{Weingarth},
	Daniel ; \textsc{Presser}, Volker  ; \textsc{Roij}, Ren{\'{e}} van:
	\newblock Heat-to-current conversion of low-grade heat from a thermocapacitive
	cycle by supercapacitors.
	\newblock {In: }\emph{Energy {\&} Environmental Science} 8 (2015), Nr. 8,
	2396--2401.
	\newblock \url{http://dx.doi.org/10.1039/c5ee01192b}. --
	\newblock DOI 10.1039/c5ee01192b
	
	\bibitem[18]{Shapiro2012Infrared}
	\textsc{Shapiro}, Mikhail~G. ; \textsc{Homma}, Kazuaki ; \textsc{Villarreal},
	Sebastian ; \textsc{Richter}, Claus-Peter  ; \textsc{Bezanilla}, Francisco:
	\newblock Infrared light excites cells by changing their electrical
	capacitance.
	\newblock {In: }\emph{Nature Communications} 3 (2012), Januar, Nr. 736.
	\newblock \url{http://dx.doi.org/10.1038/ncomms1742}. --
	\newblock DOI 10.1038/ncomms1742
	
	\bibitem[19]{Plaksin2018Thermal}
	\textsc{Plaksin}, Michael ; \textsc{Shapira}, Einat ; \textsc{Kimmel}, Eitan  ;
	\textsc{Shoham}, Shy:
	\newblock Thermal Transients Excite Neurons through Universal Intramembrane
	Mechanoelectrical Effects.
	\newblock {In: }\emph{Physical Review X} 8 (2018), M{\^^b a}rz, Nr. 1, 011043.
	\newblock \url{http://dx.doi.org/10.1103/physrevx.8.011043}. --
	\newblock DOI 10.1103/physrevx.8.011043
	
	\bibitem[20]{Lichtervelde2020heat}
	\textsc{Lichtervelde}, Aymar C.~L. ; \textsc{Souza}, J.~P.  ; \textsc{Bazant},
	Martin~Z.:
	\newblock Heat of nervous conduction: A thermodynamic framework.
	\newblock {In: }\emph{Physical Review E} 101 (2020), Februar, Nr. 2, 022406.
	\newblock \url{http://dx.doi.org/10.1103/physreve.101.022406}. --
	\newblock DOI 10.1103/physreve.101.022406
	
	\bibitem[21]{Gouy1910sur}
	\textsc{Gouy}, M.:
	\newblock Sur la constitution de la charge {\'{e}}lectrique {\`{a}} la surface
	d{\textquotesingle}un {\'{e}}lectrolyte.
	\newblock {In: }\emph{Journal de Physique Th{\'{e}}orique et Appliqu{\'{e}}e} 9
	(1910), Nr. 1, 457--468.
	\newblock \url{http://dx.doi.org/10.1051/jphystap:019100090045700}. --
	\newblock DOI 10.1051/jphystap:019100090045700
	
	\bibitem[22]{Chapman1913contribution}
	\textsc{Chapman}, David~L.:
	\newblock A contribution to the theory of electrocapillarity.
	\newblock {In: }\emph{The London, Edinburgh, and Dublin Philosophical Magazine
		and Journal of Science} 25 (1913), April, Nr. 148, 475--481.
	\newblock \url{http://dx.doi.org/10.1080/14786440408634187}. --
	\newblock DOI 10.1080/14786440408634187
	
	\bibitem[23]{Stern1924theorie}
	\textsc{Stern}, Otto:
	\newblock Zur {T}heorie der elektrolytischen {D}oppelschicht.
	\newblock {In: }\emph{Zeitschrift f{\"u}r {E}lektrochemie und angewandte
		physikalische {C}hemie} 30 (1924), Nr. 21-22, S. 508--516
	
	\bibitem[24]{Evans1979nature}
	\textsc{Evans}, R.:
	\newblock The nature of the liquid-vapour interface and other topics in the
	statistical mechanics of non-uniform, classical fluids.
	\newblock {In: }\emph{Advances in Physics} 28 (1979), April, Nr. 2, 143--200.
	\newblock \url{http://dx.doi.org/10.1080/00018737900101365}. --
	\newblock DOI 10.1080/00018737900101365
	
	\bibitem[25]{Rosenfeld1989free}
	\textsc{Rosenfeld}, Yaakov:
	\newblock Free-energy model for the inhomogeneous hard-sphere fluid mixture and
	density-functional theory of freezing.
	\newblock {In: }\emph{Physical Review Letters} 63 (1989), August, Nr. 9,
	980--983.
	\newblock \url{http://dx.doi.org/10.1103/physrevlett.63.980}. --
	\newblock DOI 10.1103/physrevlett.63.980
	
	\bibitem[26]{Roth2010fundamental}
	\textsc{Roth}, Roland:
	\newblock Fundamental measure theory for hard-sphere mixtures: a review.
	\newblock {In: }\emph{Journal of Physics: Condensed Matter} 22 (2010), Januar,
	Nr. 6, 063102.
	\newblock \url{http://dx.doi.org/10.1088/0953-8984/22/6/063102}. --
	\newblock DOI 10.1088/0953--8984/22/6/063102
	
	\bibitem[27]{Oettel2012description}
	\textsc{Oettel}, M. ; \textsc{Dorosz}, S. ; \textsc{Berghoff}, M. ;
	\textsc{Nestler}, B.  ; \textsc{Schilling}, Tanja:
	\newblock Description of hard-sphere crystals and crystal-fluid interfaces: A
	comparison between density functional approaches and a phase-field crystal
	model.
	\newblock {In: }\emph{Physical Review E} 86 (2012), August, Nr. 2, 021404.
	\newblock \url{http://dx.doi.org/10.1103/physreve.86.021404}. --
	\newblock DOI 10.1103/physreve.86.021404
	
	\bibitem[28]{Haertel2015anisotropic}
	\textsc{H\"{a}rtel}, Andreas ; \textsc{Kohl}, Matthias  ;
	\textsc{Schmiedeberg}, Michael:
	\newblock Anisotropic pair correlations in binary and multicomponent
	hard-sphere mixtures in the vicinity of a hard wall: A combined density
	functional theory and simulation study.
	\newblock {In: }\emph{Physical Review E} 92 (2015), Oktober, Nr. 4, 042310.
	\newblock \url{http://dx.doi.org/10.1103/physreve.92.042310}. --
	\newblock DOI 10.1103/physreve.92.042310
	
	\bibitem[29]{Roth2016shells}
	\textsc{Roth}, Roland ; \textsc{Gillespie}, Dirk:
	\newblock Shells of charge: a density functional theory for charged hard
	spheres.
	\newblock {In: }\emph{Journal of Physics: Condensed Matter} 28 (2016), April,
	Nr. 24, 244006.
	\newblock \url{http://dx.doi.org/10.1088/0953-8984/28/24/244006}. --
	\newblock DOI 10.1088/0953--8984/28/24/244006
	
	\bibitem[30]{Haertel2017structure}
	\textsc{H\"{a}rtel}, Andreas:
	\newblock Structure of electric double layers in capacitive systems and to what
	extent (classical) density functional theory describes it.
	\newblock {In: }\emph{Journal of Physics: Condensed Matter} 29 (2017),
	September, Nr. 42, 423002.
	\newblock \url{http://dx.doi.org/10.1088/1361-648x/aa8342}. --
	\newblock DOI 10.1088/1361--648x/aa8342
	
	\bibitem[31]{Cats2021primitive}
	\textsc{Cats}, P. ; \textsc{Evans}, R. ; \textsc{H\"{a}rtel}, A.  ;
	\textsc{Roij}, R. van:
	\newblock Primitive model electrolytes in the near and far field: Decay lengths
	from {DFT} and simulations.
	\newblock {In: }\emph{The Journal of Chemical Physics} 154 (2021), M{\^^b a}rz,
	Nr. 12, 124504.
	\newblock \url{http://dx.doi.org/10.1063/5.0039619}. --
	\newblock DOI 10.1063/5.0039619
	
	\bibitem[32]{Cats2021differential}
	\textsc{Cats}, Peter ; \textsc{Roij}, Ren{\'{e}} van:
	\newblock The differential capacitance as a probe for the electric double layer
	structure and the electrolyte bulk composition.
	\newblock {In: }\emph{The Journal of Chemical Physics} 155 (2021), September,
	Nr. 10, 104702.
	\newblock \url{http://dx.doi.org/10.1063/5.0064315}. --
	\newblock DOI 10.1063/5.0064315
	
	\bibitem[33]{Tansel2006significance}
	\textsc{Tansel}, Berrin ; \textsc{Sager}, John ; \textsc{Rector}, Tony ;
	\textsc{Garland}, Jay ; \textsc{Strayer}, Richard~F. ; \textsc{Levine},
	Lanfang ; \textsc{Roberts}, Michael ; \textsc{Hummerick}, Mary  ;
	\textsc{Bauer}, Jan:
	\newblock Significance of hydrated radius and hydration shells on ionic
	permeability during nanofiltration in dead end and cross flow modes.
	\newblock {In: }\emph{Separation and Purification Technology} 51 (2006),
	August, Nr. 1, 40--47.
	\newblock \url{http://dx.doi.org/10.1016/j.seppur.2005.12.020}. --
	\newblock DOI 10.1016/j.seppur.2005.12.020
	
	\bibitem[34]{Lyashchenko2010dielectric}
	\textsc{Lyashchenko}, Andrey ; \textsc{Lileev}, Alexander:
	\newblock Dielectric Relaxation of Water in Hydration Shells of Ions.
	\newblock {In: }\emph{Journal of Chemical {\&} Engineering Data} 55 (2010),
	April, Nr. 5, 2008--2016.
	\newblock \url{http://dx.doi.org/10.1021/je900961m}. --
	\newblock DOI 10.1021/je900961m
	
	\bibitem[35]{Bankura2013hydration}
	\textsc{Bankura}, Arindam ; \textsc{Carnevale}, Vincenzo  ; \textsc{Klein},
	Michael~L.:
	\newblock Hydration structure of salt solutions fromab initiomolecular
	dynamics.
	\newblock {In: }\emph{The Journal of Chemical Physics} 138 (2013), Januar, Nr.
	1, 014501.
	\newblock \url{http://dx.doi.org/10.1063/1.4772761}. --
	\newblock DOI 10.1063/1.4772761
	
	\bibitem[36]{nightingale1959phenomenological}
	\textsc{{Nightingale, Jr.}}, E.~R.:
	\newblock Phenomenological Theory of Ion Solvation. {E}ffective Radii of
	Hydrated Ions.
	\newblock {In: }\emph{The Journal of Physical Chemistry} 63 (1959), September,
	Nr. 9, 1381--1387.
	\newblock \url{http://dx.doi.org/10.1021/j150579a011}. --
	\newblock DOI 10.1021/j150579a011
	
	\bibitem[37]{henderson2005monte}
	\textsc{Henderson}, Douglas ; \textsc{Gillespie}, Dirk ; \textsc{Nagy},
	T{\'{I}}mea  ; \textsc{Boda}, Dezs\"{o}:
	\newblock Monte Carlo simulation of the electric double layer: dielectric
	boundaries and the effects of induced charge.
	\newblock {In: }\emph{Molecular Physics} 103 (2005), November, Nr. 21-23,
	2851--2861.
	\newblock \url{http://dx.doi.org/10.1080/00268970500108668}. --
	\newblock DOI 10.1080/00268970500108668
	
	\bibitem[38]{buyukdagli2012dipolar}
	\textsc{Buyukdagli}, Sahin ; \textsc{Ala-Nissila}, Tapio:
	\newblock Dipolar depletion effect on the differential capacitance of
	carbon-based materials.
	\newblock {In: }\emph{{EPL} (Europhysics Letters)} 98 (2012), Juni, Nr. 6,
	60003.
	\newblock \url{http://dx.doi.org/10.1209/0295-5075/98/60003}. --
	\newblock DOI 10.1209/0295--5075/98/60003
	
	\bibitem[39]{buyukdagli2012excluded}
	\textsc{Buyukdagli}, Sahin ; \textsc{Ala-Nissila}, Tapio:
	\newblock Excluded volume effects in macromolecular forces and ion-interface
	interactions.
	\newblock {In: }\emph{The Journal of Chemical Physics} 136 (2012), Februar, Nr.
	7, 074901.
	\newblock \url{http://dx.doi.org/10.1063/1.3684880}. --
	\newblock DOI 10.1063/1.3684880
	
	\bibitem[40]{valleau1982electrical}
	\textsc{Valleau}, J.~P. ; \textsc{Torrie}, G.~M.:
	\newblock The electrical double layer. {III}. {M}odified {G}ouy-{C}hapman
	theory with unequal ion sizes.
	\newblock {In: }\emph{The Journal of Chemical Physics} 76 (1982), Mai, Nr. 9,
	4623--4630.
	\newblock \url{http://dx.doi.org/10.1063/1.443542}. --
	\newblock DOI 10.1063/1.443542
	
	\bibitem[41]{yu2006effects}
	\textsc{Yu}, Jiang ; \textsc{Aguilar-Pineda}, Gabriel~E. ;
	\textsc{Antill{\'{o}}n}, A. ; \textsc{Dong}, Shi-Hai  ;
	\textsc{Lozada-Cassou}, M.:
	\newblock The effects of unequal ionic sizes for an electrolyte in a charged
	slit.
	\newblock {In: }\emph{Journal of Colloid and Interface Science} 295 (2006),
	M{\^^b a}rz, Nr. 1, 124--134.
	\newblock \url{http://dx.doi.org/10.1016/j.jcis.2005.08.016}. --
	\newblock DOI 10.1016/j.jcis.2005.08.016
	
	\bibitem[42]{Hoover1968melting}
	\textsc{Hoover}, William~G. ; \textsc{Ree}, Francis~H.:
	\newblock Melting Transition and Communal Entropy for Hard Spheres.
	\newblock {In: }\emph{The Journal of Chemical Physics} 49 (1968), Oktober, Nr.
	8, 3609--3617.
	\newblock \url{http://dx.doi.org/10.1063/1.1670641}. --
	\newblock DOI 10.1063/1.1670641
	
	\bibitem[43]{Oettel2010free}
	\textsc{Oettel}, M. ; \textsc{G\"{o}rig}, S. ; \textsc{H\"{a}rtel}, A. ;
	\textsc{L\"{o}wen}, H. ; \textsc{Radu}, M.  ; \textsc{Schilling}, T.:
	\newblock Free energies, vacancy concentrations, and density distribution
	anisotropies in hard-sphere crystals: A combined density functional and
	simulation study.
	\newblock {In: }\emph{Physical Review E} 82 (2010), November, Nr. 5, 051404.
	\newblock \url{http://dx.doi.org/10.1103/physreve.82.051404}. --
	\newblock DOI 10.1103/physreve.82.051404
	
	\bibitem[44]{Hohenberg1964inhomogeneous}
	\textsc{Hohenberg}, P. ; \textsc{Kohn}, W.:
	\newblock Inhomogeneous Electron Gas.
	\newblock {In: }\emph{Physical Review} 136 (1964), November, Nr. 3B,
	B864--B871.
	\newblock \url{http://dx.doi.org/10.1103/physrev.136.b864}. --
	\newblock DOI 10.1103/physrev.136.b864
	
	\bibitem[45]{Mermin1965thermal}
	\textsc{Mermin}, N.~D.:
	\newblock Thermal Properties of the Inhomogeneous Electron Gas.
	\newblock {In: }\emph{Physical Review} 137 (1965), M{\^^b a}rz, Nr. 5A,
	A1441--A1443.
	\newblock \url{http://dx.doi.org/10.1103/physrev.137.a1441}. --
	\newblock DOI 10.1103/physrev.137.a1441
	
	\bibitem[46]{Hansen2013Theory}
	\textsc{Hansen}, Jean-Pierre ; \textsc{McDonald}, Ian~R.:
	\newblock \emph{Theory of simple liquids}.
	\newblock 4.
	\newblock Elsevier, 2013
	
	\bibitem[47]{HansenGoos2006density}
	\textsc{Hansen-Goos}, Hendrik ; \textsc{Roth}, Roland:
	\newblock Density functional theory for hard-sphere mixtures: the White Bear
	version mark {II}.
	\newblock {In: }\emph{Journal of Physics: Condensed Matter} 18 (2006), August,
	Nr. 37, 8413--8425.
	\newblock \url{http://dx.doi.org/10.1088/0953-8984/18/37/002}. --
	\newblock DOI 10.1088/0953--8984/18/37/002
	
	\bibitem[48]{Tarazona2000density}
	\textsc{Tarazona}, P.:
	\newblock Density Functional for Hard Sphere Crystals: A Fundamental Measure
	Approach.
	\newblock {In: }\emph{Physical Review Letters} 84 (2000), Januar, Nr. 4,
	694--697.
	\newblock \url{http://dx.doi.org/10.1103/physrevlett.84.694}. --
	\newblock DOI 10.1103/physrevlett.84.694
	
	\bibitem[49]{Ng1974Hypernetted}
	\textsc{Ng}, Kin-Chue:
	\newblock Hypernetted chain solutions for the classical one-component plasma up
	to {$\Gamma$}=7000.
	\newblock {In: }\emph{The Journal of Chemical Physics} 61 (1974), Oktober, Nr.
	7, 2680--2689.
	\newblock \url{http://dx.doi.org/10.1063/1.1682399}. --
	\newblock DOI 10.1063/1.1682399
	
	\bibitem[50]{breitsprecher2017effect}
	\textsc{Breitsprecher}, Konrad ; \textsc{Abele}, Manuel ; \textsc{Kondrat},
	Svyatoslav  ; \textsc{Holm}, Christian:
	\newblock The effect of finite pore length on ion structure and charging.
	\newblock {In: }\emph{The Journal of Chemical Physics} 147 (2017), September,
	Nr. 10, 104708.
	\newblock \url{http://dx.doi.org/10.1063/1.4986346}. --
	\newblock DOI 10.1063/1.4986346
	
	\bibitem[51]{breitsprecher2018charge}
	\textsc{Breitsprecher}, Konrad ; \textsc{Holm}, Christian  ; \textsc{Kondrat},
	Svyatoslav:
	\newblock Charge Me Slowly, {I} Am in a Hurry: Optimizing
	Charge{\textendash}Discharge Cycles in Nanoporous Supercapacitors.
	\newblock {In: }\emph{{ACS} Nano} 12 (2018), August, Nr. 10, 9733--9741.
	\newblock \url{http://dx.doi.org/10.1021/acsnano.8b04785}. --
	\newblock DOI 10.1021/acsnano.8b04785
	
	\bibitem[52]{Marcus1991thermodynamics}
	\textsc{Marcus}, Yizhak:
	\newblock Thermodynamics of solvation of ions. Part 5.{\textemdash}{G}ibbs free
	energy of hydration at 298.15 {K}.
	\newblock {In: }\emph{Journal of the Chemical Society, Faraday Transactions} 87
	(1991), Nr. 18, 2995--2999.
	\newblock \url{http://dx.doi.org/10.1039/ft9918702995}. --
	\newblock DOI 10.1039/ft9918702995
	
	\bibitem[53]{Lindner2019Entropy}
	\textsc{Lindner}, Jeannette ; \textsc{Weick}, Fabian ; \textsc{Endres}, Frank
	; \textsc{Schuster}, Rolf:
	\newblock Entropy Changes upon Double Layer Charging at a (111)-Textured Au
	Film in Pure 1-Butyl-1-Methylpyrrolidinium
	Bis[(trifluoromethyl)sulfonyl]imide Ionic Liquid.
	\newblock {In: }\emph{The Journal of Physical Chemistry C} 124 (2019),
	Dezember, Nr. 1, 693--700.
	\newblock \url{http://dx.doi.org/10.1021/acs.jpcc.9b09871}. --
	\newblock DOI 10.1021/acs.jpcc.9b09871
	
\end{thebibliography}
\end{document}